\documentclass{article}

\usepackage{PRIMEarxiv}

\usepackage[utf8]{inputenc} 
\usepackage[T1]{fontenc}    
\usepackage{hyperref}       
\usepackage{url}            
\usepackage{booktabs}       
\usepackage{amsfonts}       
\usepackage{nicefrac}       
\usepackage{microtype}      
\usepackage{lipsum}
\usepackage{fancyhdr}       
\usepackage{graphicx}       
\graphicspath{{media/}}     

\usepackage[colorinlistoftodos]{todonotes}
\usepackage{xcolor}
\usepackage{soul}
\usepackage{multirow}
\usepackage{caption}
\usepackage{adjustbox}
\usepackage{array}
\usepackage{longtable}
\usepackage{amsmath}
\usepackage{amsfonts}
\usepackage{amssymb}
\DeclareMathOperator*{\argmax}{arg\,max}
\DeclareMathOperator*{\argmin}{arg\,min}

\title{A Comprehensive Survey on Context-Aware Multi-Agent
Systems: Techniques, Applications, Challenges and Future
Directions
}

\author{
  Hung Du, Srikanth Thudumu, Hy Nguyen, Rajesh Vasa, Kon Mouzakis \\
  Applied Artificial Intelligence Institute ($A^2I^2$) \\
  Deakin University, Geelong, VIC, Australia, 3216 \\
  \texttt{\{hung.du, srikanth.thudumu, hy.nguyen, rajesh.vasa, kon.mouzakis\}@deakin.edu.au} \\
}

\begin{document}
\maketitle

\newcommand{\note}[1]{\textcolor{black}{[NOTE: #1]}}
\newcommand{\rnote}[1]{\textcolor{red}{[NOTE: #1]}}
\newcommand{\blue}[1]{\textcolor{black}{#1}}
\newcolumntype{L}[1]{>{\raggedright\arraybackslash}p{#1}}
\newcolumntype{C}[1]{>{\centering\arraybackslash\hspace{0pt}}p{#1}}

\begin{abstract}
Research interest in agents is rising, particularly in Artificial Intelligence (AI) techniques such as deep multi-modal representation learning, deep graph learning, and deep reinforcement learning, which suggest potential for human-like task performance within a multi-agent framework. However, challenges exist in enabling these agents to coordinate, learn, reason, predict, and navigate uncertainties in dynamic environments. Context awareness is crucial for enhancing multi-agent systems in such situations. Current research often addresses context-awareness or multi-agent system issues separately, leaving a gap in comprehensive surveys that explore both fields, especially regarding the five essential agent capabilities (Sense-Learn-Reason-Predict-Act). This survey fills that gap by presenting a unified architecture and taxonomy for developing Context-Aware Multi-Agent Systems (CA-MAS), which are vital for improving agent robustness and adaptability in real-world environments. We provide a comprehensive overview of state-of-the-art context-awareness and multi-agent systems, followed by a framework for CA-MAS development. We detail the properties of context-awareness and multi-agent systems, and a general process for context-aware systems, highlighting approaches from various domains such as ambient intelligence, autonomous navigation, digital assistance, disaster relief management, education, energy efficiency \& sustainability, IoT, and supply chain management. Finally, we discuss existing CA-MAS challenges and propose future research directions.
\end{abstract}


\section{Introduction}
\blue{In recent years, the Artificial Intelligence (AI) community has been approaching a paradigm shift, transitioning from creating AI systems for passive, structured tasks to developing AI systems that are capable of dynamic, agentic roles in diverse and complex environments. Additionally, developing autonomous agents with human-like task performance is progressing rapidly \cite{bubeck2023sparks,hauptman2023adapt,wang2024survey} due to the following breakthroughs:
\begin{enumerate}
    \item \textbf{An agent can preceive contextual information:} The development of Context-Aware Systems (CAS) illustrates the agent's ability to sense and model contextual information and utilizes such data for learning and reasoning purposes \cite{baldauf2007survey,hoareau2009modeling,lee2011survey,perera2013context,abbas2015survey,alegre2016engineering,pradeep2019mom,kulkarni2020context,islam2021context,vahdat2021survey,casillo2023context} (see also Section \ref{sec:cas}).
    \item \textbf{An agent can learn any sensory data:} Advancements in deep learning techniques show that multiple types of sensory data, including images, audio, text, and signals, can be transformed into machine-readable formats \cite{lecun2015deep,pouyanfar2018survey}. These formats are subsequently encoded into learnable tokens through the use of Transformer architectures \cite{vaswani2017attention,lin2022survey}. Frameworks, such as joint representation, coordinated representation, and Encoder-Decoder facilitate the conversion between tokens of different data modalities \cite{guo2019deep,baltruvsaitis2018multimodal,xu2023multimodal}. These innovations enable an agent to represent and work with various type of sensory data.
    \item \textbf{An agent can reason:} The semantic relationships of sensory data are encapsulated in the form of an ontology \cite{moller2008ontology,asim2019use,ji2021survey}. An agent can utilize these relationships for reasoning using rule-based, case-based, or graph-based approaches (refer to Section \ref{sec:camas_reason}). Moreover, advancements in Deep Reinforcement Learning (DRL) techniques enable the agent to plan and reason with goals (refer to Section \ref{sec:camas_reason}). Additionally, the representation of the ontology, generated by Graph Neural Networks (GNN) or their variants \cite{wu2020comprehensive,zhang2020deep}, can be integrated with DRL to further enhance the agent's reasoning capabilities.
    \item \textbf{An agent can predict and take action:} The agent estimates and predicts future states based on its model of historical data and observations. This prediction process can be carried out using weight schemes, probabilistic models, or reward-based models (see also Section \ref{sec:camas_predict}). Additionally, DRL techniques demonstrate that the agent can choose optimal actions in various scenarios by interactively imitating the behaviors of other agents and humans \cite{mnih2015human,christiano2017deep,wirth2017survey,hussein2017imitation,jara2019theory,rafailov2024direct}.
    \item \textbf{An agent can coordinate with other agents:} The development of Multi-Agent Systems (MAS) demonstrates the agent's capability of communicating and coordinating with other agents to achieve a goal within some set of constraints \cite{horling2004survey,qin2016recent,dorri2018multi,liu2018survey,li2019survey,litimein2021survey,amirkhani2022consensus,liu2023survey,zhu2024survey} (see also Section \ref{sec:mas}).
\end{enumerate}
To understand dynamic and complex environments, an autonomous agent must possess a combination of five capabilities: Sense, Learn, Reason, Predict, and Act. For example, the agent learns and reasons about the behaviors, goals, and beliefs of other agents or humans. This understanding enables the agent to predict and act within the dynamic environment through self-evaluation \cite{hong2023metagpt,fernando2023promptbreeder,madaan2024self}. Note that
}
the limitations of a single autonomous agent such as inefficiency, high cost, and unreliability \cite{li2019survey,amirkhani2022consensus} become apparent when addressing distributed complex tasks such as microgrid control \cite{kantamneni2015survey}, resource allocation in cloud computing \cite{de2019survey}, as well as in the realms of computer networking and security, and other complex tasks \cite{dorri2018multi}. Overcoming these limitations requires the implementation of a system that allocates task responsibilities among autonomous agents, enabling effective coordination and communication during task resolution. Additionally, in a specific task, a collective of autonomous agents needs to adapt to changes of the environment by constantly learning and updating their knowledge over time. This results in the growing attention to the field of context-aware multi-agent systems where agents comprehend their knowledge according to perceived contextual information to adapt to any situation and optimally solve allocated tasks \blue{and achieve the global goal}. Application domains of such systems include \blue{autonomous navigation, ambient intelligence, supply chain management, Internet of Things (IoT), disaster relief management, energy efficiency \& sustainability, digital assistance and education, and other complex problems (see also Section \ref{sec:ca_mas}).}

\begin{figure*}
   \includegraphics[width=\linewidth]{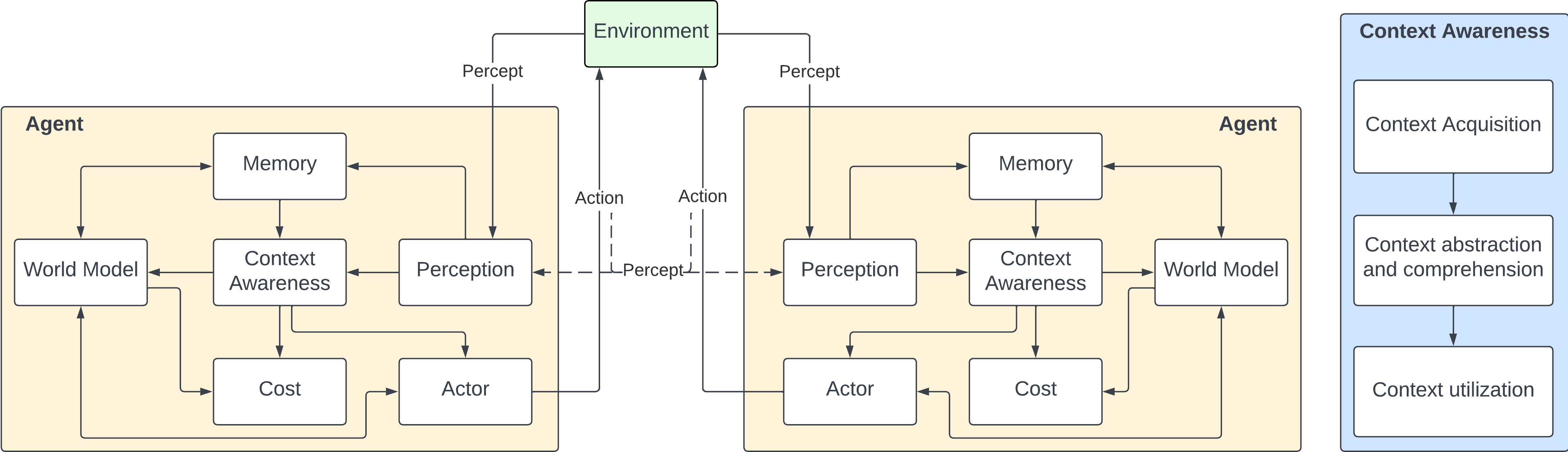}
   \caption{An Overview of Context-Aware Multi-Agent Systems}
   \label{fig:overview}
\end{figure*}

Multi-Agent Systems (MAS) comprise multiple autonomous agents that interact within a shared environment, autonomously making decisions to accomplish tasks or address complex problems. An autonomous agent in MAS is endowed with initial knowledge about a given task and possesses its own set of goals. While engaged in task-solving, the agent interacts with other agents or the environment to perceive and comprehend information. It independently makes decisions based on its objectives, existing knowledge, and observations, subsequently executing actions \cite{zhu2024survey,zhou2023multi}. Depending on the task's characteristics, agents can collaborate or compete strategically to outperform opponents. These attributes confer flexibility and adaptability on agents in dynamic environments, making MAS suited for addressing complex problems. \blue{Furthermore}, an autonomous agent forms its beliefs based on its knowledge and observations of the environment and other agents. Moreover, motivations for actions can vary among agents. Consequently, determining the optimal solution for a specific task necessitates effective communication, coordination, or competition in MAS. In cooperative settings, achieving consensus, and a shared agreement on a particular interest is imperative for autonomous agents. Conversely, in competitive scenarios, agents must analyze the behavior of opponents, anticipate negative outcomes, and devise strategies to address such challenges. It is worth noting that attaining consensus or understanding the behavior of other agents requires an autonomous agent to be aware of the context, including roles, organizational structure, situations, and location \cite{horling2004survey,amirkhani2022consensus}. Therefore, the integration of context-aware systems into multi-agent systems, as known as context-aware multi-agent systems, becomes essential.

Context-Aware Systems (CAS) pertain to systems that dynamically adapt to the environment by leveraging context to retrieve prominent information for tasks or problems. Context encompasses various perceptual elements such as people, location, physical or virtual objects, events, and other information delineating the situation of an autonomous agent in a specific environment \cite{abowd1999towards}. The CAS process involves three primary stages: context acquisition, context abstraction and comprehension, and context utilization. \cite{lee2011survey,perera2013context,alegre2016engineering}. Through this continuous process, autonomous agents acquire contextual cues to comprehend the current environmental state and undertake actions relevant to the situation. In the context of autonomous agents, contextual information encompasses task objectives, organizational structure, agent roles, and temporal aspects. Such context assists agents precisely retrieving relevant information for task accomplishment \cite{hong2023metagpt,qian2023communicative,chen2023agentverse}.

Considerable efforts have focused on either MAS or CAS. For instance, \blue{Dorri \textit{et al.}} \cite{dorri2018multi} conducted a survey outlining aspects of MAS, such as agent properties, organizational structures, coordination control, and communication. The survey also emphasized distinctions between MAS and related systems, such as object-oriented programming and expert systems (see Section \ref{sec:mas}). Other surveys have delved into specific aspects of MAS, including consensus \cite{qin2016recent,li2019survey,amirkhani2022consensus}, communication \cite{zhu2024survey}, formation control \cite{liu2018survey,litimein2021survey,liu2023survey}, organizational structures \cite{horling2004survey}, and deep reinforcement learning (DRL) for MAS \blue{\cite{busoniu2008comprehensive,hernandez2019survey,nguyen2020deep,zhang2021multi,gronauer2022multi,zhou2023multi}}. On the other hand, surveys related to CAS can be categorized into two streams: fundamental overviews and domain-specific applications. Fundamental overviews highlight the general processes of CAS, context modeling approaches, and design principles for architecting CAS \cite{baldauf2007survey,hoareau2009modeling,lee2011survey,alegre2016engineering,pradeep2019mom}. Research on domain-specific applications of CAS includes recommendation systems \cite{abbas2015survey,kulkarni2020context,casillo2023context}, Internet of Thing (IoT) \cite{perera2013context}, cloud and fog computing \cite{islam2021context}, and healthcare systems \cite{vahdat2021survey}. Despite the popularity of CAS and MAS, as well as the substantial research efforts in these fields, \blue{previous works have not considered all five capabilities of agents or framed them within the perspective of shared context awareness to achieve robustness and adaptability in dynamic environments. Therefore, there is a need for a coherent architecture, an organized taxonomy, and associated principles to build Context-Aware Multi-Agent Systems (CA-MAS). In this paper, we review prior research that connects the five capabilities of agents and present a cohesive framework for constructing CA-MAS. Additionally, we highlight the existing challenges of CA-MAS and propose future directions to address them.
}

The rest of this paper is organized as follows. Section \ref{sec:mas} introduces the background of MAS. Section \ref{sec:cas} highlights the definition of context, the general process of CAS, and the characteristics of CAS. Section \ref{sec:ca_mas} presents the definition of awareness in MAS and provides a thorough overview of the general process of CA-MAS. Finally, we highlight existing challenges in CA-MAS, along with future directions, and conclude this paper in Sections \ref{sec:future} and \ref{sec:conclusion}, respectively.

\section{Multi-Agent Systems} \label{sec:mas}
An agent is an autonomous and computational entity (e.g., a software, a hardware component, or a combination of both) that operates within an environment to achieve specific tasks or goals \cite{nguyen2020deep,gronauer2022multi}. Furthermore, an agent is characterized by its ability to perceive its surroundings, make decisions based on available information, and execute actions to influence the environment. Such characteristics enable the sociability, autonomy, and adaptability of an agent in MAS. As highlighted by \cite{lecun2022path}, intelligent properties of an autonomous agent encompass the following interconnected components:
\blue{
\begin{enumerate}
    \item \textbf{Perception}: The agent senses information and estimates the current environment state. Such information includes the visual, auditory, and other sensory input. For example, visual cues from images or videos allow the agent to plan pedestrians' motion \cite{everett2018motion} or traffic scenarios \cite{chen2020delay,xie2021congestion} to avoid collision. Furthermore, Oliveira \textit{et al.} \cite{oliveira2022multi} proposed a MAS that proceeds user interaction through audio and gestures to understand and addresses specific needs of elderly people. Moreover, acoustics, Radio Frequency (RF), optics, and magnetics are four general types of sensory data for underwater multi-agent systems \cite{zhou2021survey}.
    \item \textbf{Memory}: The agent stores past, present, and predicted future states of the environment along with their intrinsic costs. Such information in the memory is accessible and modifiable by the world model to make temporal decisions or correct inconsistent and missing information about the current states of the world. On the other hand, the critic module learns from historical information to estimate their costs, supporting the agent in selecting a set of actions to optimally achieve its goal. Due to its limited memory capacity, the agent may not retain information for long periods. As it explores the environment, some data in its memory inevitably fades. Therefore, the experience replay mechanism proposed in \cite{lin1992reinforcement} aims to assist the agent in re-learning forgotten information. However, not all forgotten information is necessarily worth re-learning, as it may not be immediately useful to the agent. Hence, Schaul \textit{et al.} \cite{DBLP:journals/corr/SchaulQAS15} proposed the prioritized experience replay method, which prioritizes information that frequently occurred with high rewards during the learning process.
    \item \textbf{World Model}: The agent utilizes its current understanding of the world and past experiences to predict missing information and future states. Using the world model, the agent can anticipate the effects and outcomes of various actions to mitigate risks while searching for optimal strategies that include good actions and policies (see also Sections \ref{sec:camas_reason} and \ref{sec:camas_predict}). However, it is important to acknowledge that gathering enough data to fully capture the environment's behavior may not always be practical. Therefore, the predictions made by the world model could be inaccurate or uncertain.
    \item \textbf{Actor}: The agent determines an optimal sequence of actions from proposed action plans to optimize the estimated future cost. After completing the optimization process, the agent selects an action from that sequence at a given time. Depending on the context, actions can be categorized into two types: discrete actions (e.g., controlling traffic lights, recommending from a finite set of items, or allocating energy to different areas) and continuous actions (e.g., physical control tasks, or adjusting heating, ventilation, and air conditioning settings in buildings based on weather conditions).
    \item \textbf{Cost}: The agent assesses and evaluates its actions and knowledge based on the measured cost, which comprises two components: the intrinsic cost and the critic. The intrinsic cost is predetermined by the agent's behavior and remains constant throughout the learning process, providing insight into the nature of the agent's behavior at different times. On the other hand, the critic is trainable during the learning process and aims to estimate long-term outcomes based on the current state of the world and the agent's past experiences. An example of the critic component can be found in Actor-Critic methods \cite{prokhorov1997adaptive,sutton1999policy,peters2008natural,silver2014deterministic,mnih2016asynchronous,DBLP:journals/corr/LillicrapHPHETS15,haarnoja2018soft}. These methods generally involve two stages: (1) the agent selects an action from the optimal action sequence using the \textit{Actor} component; and (2) the \textit{Critic} component estimates the cost of that action by approximating the expected future values of the reward. The agent's overall objective is to take actions that minimize the average cost of the intrinsic cost and the critic.
    \item \textbf{Configurator (also known as the Context-Awareness component)}: Depending on the situation presented by the environment, the agent adjusts configurations across its components to adapt to changes and progress towards its goals. For example, during the sensing process within the \textit{Perception} component, the agent filters out irrelevant information for a given task. Additionally, the functions of the \textit{Critic} component, learning weights of the world model, and policies can be dynamically configured to guide the agent towards specific sub-goals within a larger objective. While not all components are configurable in real-time, two strategies have been explored in the literature: online learning strategies and offline learning strategies \cite{sutton2018reinforcement,levine2020offline,shakya2023reinforcement}.
\end{enumerate}
}
The environment in a MAS serves as the operational space where agents operate and pursue goals. It consists of perceptual data and dynamic changes of state \cite{dorri2018multi}. This environment provides the context guiding an agent's decision-making process. As illustrated in Figure \ref{fig:overview}, the task-solving process involves seven steps \cite{mostafa2013formulating,dorri2018multi}: perceiving information from the environment, storing it in memory, processing the information based on the task, comprehending task requirements, formulating a plan, executing actions, and learning from outcomes.

Multi-Agent Systems (MAS) comprise multiple autonomous agents interacting with each other and their environment to achieve goals. Four main advantages of using a MAS over a single-agent one include cost-effectiveness, reliability, scalability, and robustness in handling complex tasks \cite{li2019survey,amirkhani2022consensus}. However, three main aspects of a MAS such as organizational structure, consensus, and formation control result in the complexity of MAS. Furthermore, three main features such as autonomy, communication, and society are three main features distinguishing between MAS, expert systems, and object-oriented programming \blue{(see also Table \ref{tab:diff_sys})} \cite{dorri2018multi}. For instance, each agent in MAS has its own goal(s) and can communicate with other agents in the network to make decisions and act on the environment, whereas objects in object-oriented programming or experts in an expert system are restricted by a set of pre-defined constraints of a particular problem. \blue{Depending on both organizational goals and agent-specific goals, relationships among agents} are formed into one of the following categories: cooperation or competition. Cooperative MAS encompasses systems where agents collaborate with a shared objective of accomplishing common goals or addressing complex problems. Achieving this purpose necessitates organizing agents into groups with coordination, establishing communication protocols among them, and reaching consensus agreements \cite{torreno2017cooperative,ismail2018survey,amirkhani2022consensus}. On the other hand, competitive MAS characterizes systems in which agents operate with the primary goal of optimizing their individual objectives. To attain this, an agent engages in strategic decision-making, with the aim of outperforming or gaining advantages over other agents \cite{liu2022overview}.
\begingroup
\footnotesize
\begin{table}[htbp]
    \centering
    \caption{The difference between object-oriented programming, expert systems and multi-agent systems \cite{dorri2018multi}}
    \label{tab:diff_sys}
    \begin{tabular}{|L{0.13\linewidth}|L{0.26\linewidth}|L{0.26\linewidth}|L{0.26\linewidth}|}
        \hline
        \textbf{Characteristics} & \textbf{Object-oriented progrmaming} & \textbf{Expert Systems} & \textbf{Multi-agent systems} \\
        \hline
        Structure & Classes and objects & Knowledge base and inference engine & Agents and their interactions \\
        \hline
        Purpose & Modeling entities and behaviors & Emulating human expert decision-making & Autonomous, coordinated problem-solving \\
        \hline
        Communication & Method calls and message passing & Rule-based or logic-based reasoning & Interaction protocols and negotiation \\
        \hline
        Learning ability & None: Static code & 	Low: Limited learning modules & High: Agents can learn and adapt autonomously \\
        \hline
        Environment & Static, predefined environments & Static, predefined environments & Dynamic, open environments \\
        \hline
    \end{tabular}
\end{table}
\endgroup

\begin{figure*}
   \includegraphics[width=0.87\linewidth]{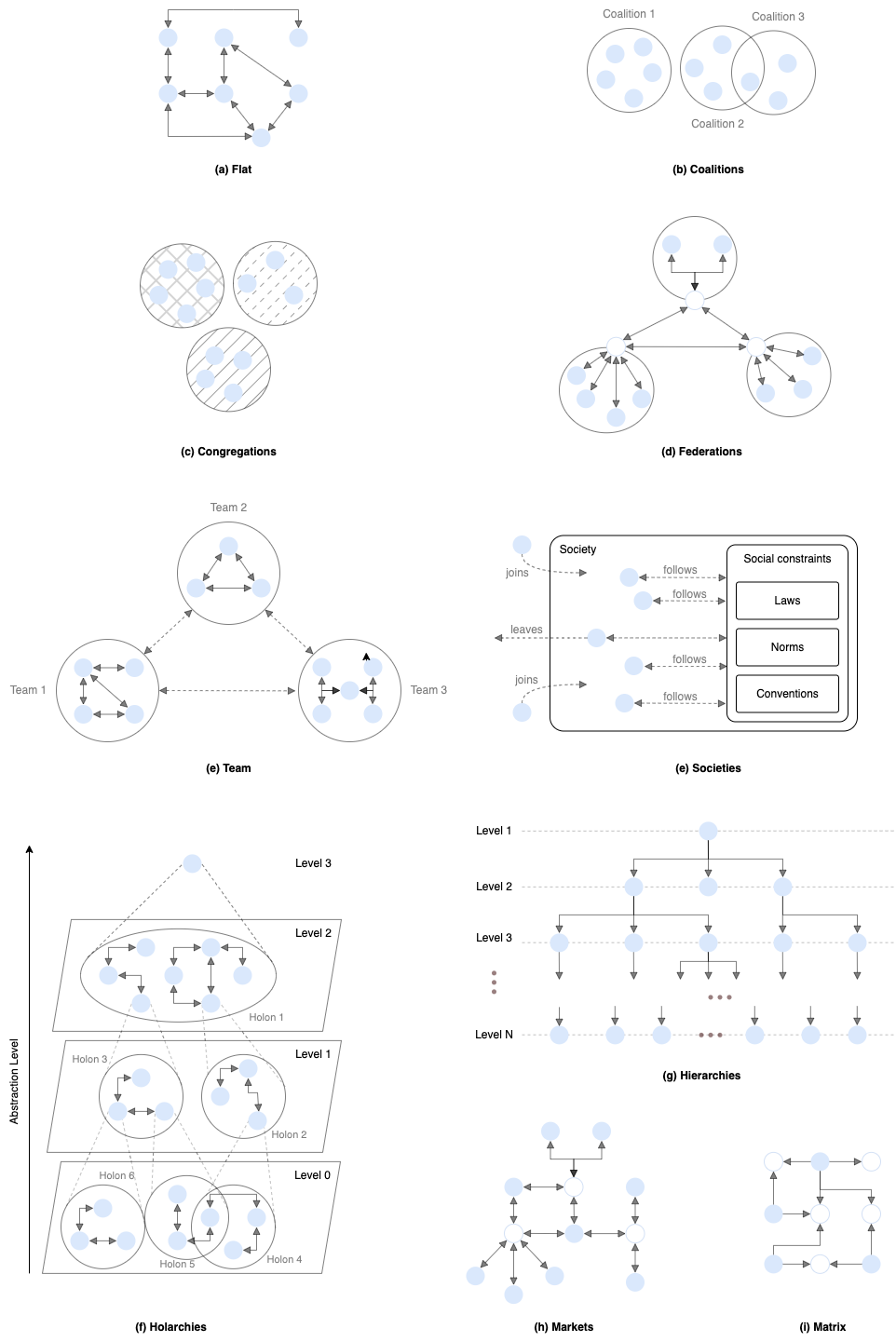}
   \captionof{figure}{Ten organizational structures of multi-agent systems. A blue circle symbolizes an agent, with its role depending on the organizational structure. For instance, in the federation structure (d), agents fulfill two distinct roles: (1) normal agents depicted as blue circles, and (2) federated agents shown as outlined circles. Similarly, in the market structure (h), there are two roles: buyers represented by blue circles and sellers by outlined circles. In the matrix structure (i), managers are represented as blue circles while workers are shown as outlined circles.}
   \label{fig:mas_org}
\end{figure*}

\begin{figure*}
   \includegraphics[width=0.95\linewidth]{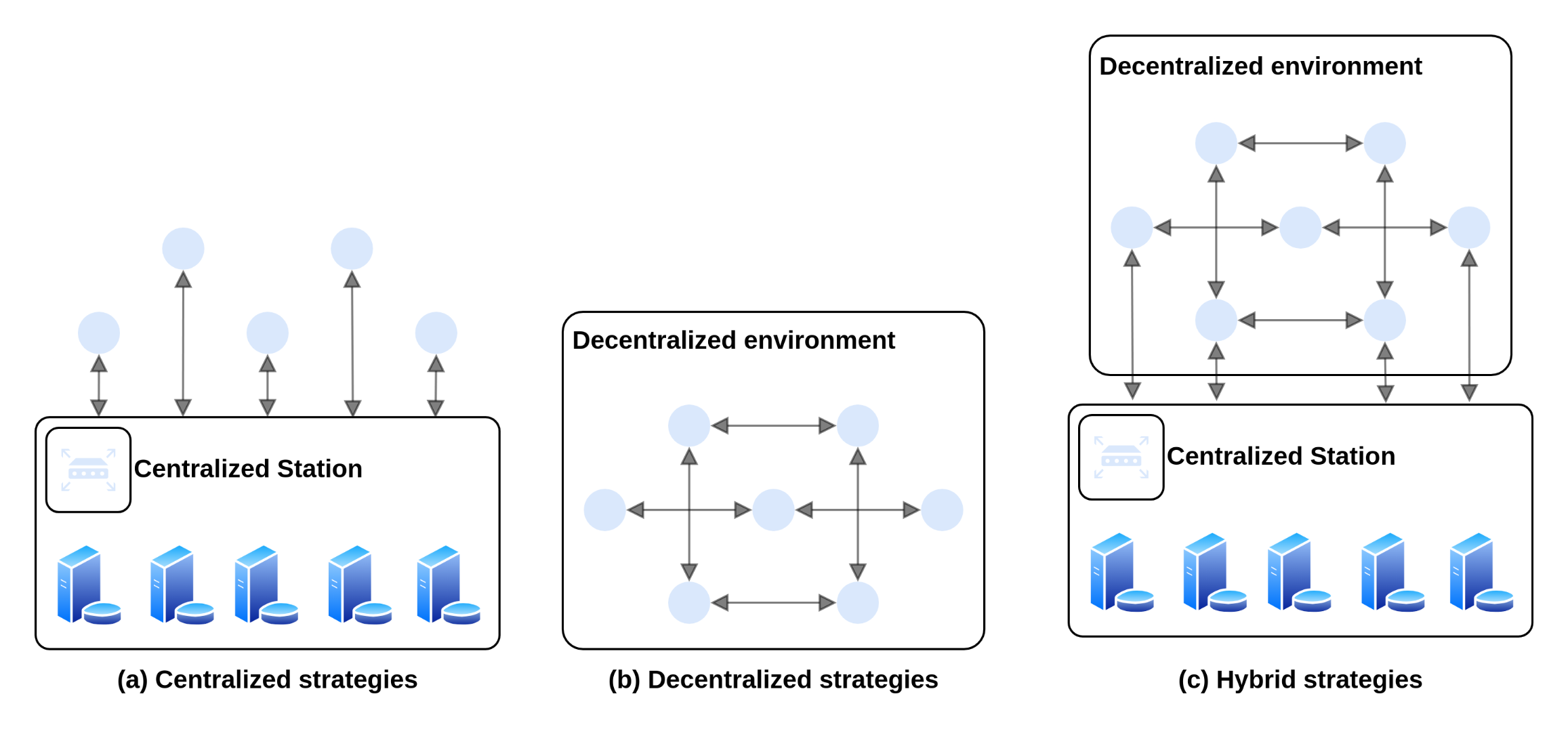}
   \captionof{figure}{Coordination strategies between agents. A blue circle represents an agent.}
   \label{fig:formation_mas}
\end{figure*}

An organizational structure, encompassing roles, relationships, and authority, plays a pivotal role in simplifying agents' models, controlling agents' behavior, and reducing uncertainty. Additionally, it facilitates the tracking and establishment of effective communication among agents when solving complex tasks. \blue{As shown in Figure \ref{fig:mas_org}}, ten organizational structures, such as (1) flat, (2) hierarchies, (3) holarchies, (4) coalitions, (5) teams, (6) congregations, (7) societies, (8) federations, (9) markets, and (10) matrix, have been identified \cite{horling2004survey}. \blue{The selection of an organizational structure depends on characteristics and agents' capabilities (see also Table \ref{tab:mas_org}). Additionally, the combination of multiple organizational structures has been applied to existing multi-agent systems \cite{horling2004survey,dastani2005coordination}. Aside from organizational structures, three mechanisms that agents form an organization include dynamic reorganization \cite{decker1993one,tambe1997towards,guessoum2003dynamic,dignum2004dynamic,picard2009reorganisation}, self-organization \cite{bernon2003adelfe,karuna2005emergent,serugendo2006self,picard2009reorganisation}, or emergence \cite{walker1995understanding,karuna2005emergent,serugendo2006self,dessalles2006emergence,teo2013formalization}. The selection of each mechanism depends on the following factors: (1) system demand, such as organizational goals and agent-specific goals; (2) external demand, such as changes of both environment and human intervention; (3) agents' awareness of the existence of organization; and (4) complex and dynamic situations yield by the environment \cite{picard2009reorganisation}. Once an organization has been formed, the coordination of agents can be operated through decentralized strategies, centralized strategies or hybrid strategies (see also Table \ref{fig:formation_mas}) \cite{hale2015cloud,do2021formation,chen2023survey,liu2023survey}. The selection of each of these strategies depends on four factors, such as real-time performance, scalability of system structure, robustness of decision-making, and solution optimality \cite{hale2015cloud,liu2023survey}.}
The diversity of agents, inferred from differences in knowledge, capability, characteristics, behaviors, and other contextual information, represents a crucial dimension of agent organization. \cite{esmaeili2016impact} demonstrated the impact of diversity in agents' capabilities and behaviors on a holonic organization in MAS. Moreover, \cite{li2021celebrating} investigated a technique focusing on behavioral diversity in MAS, aiming to control the trade-off between the heterogeneous behaviors of an agent resulting from knowledge exploration and the homogeneous behaviors obtained through knowledge sharing in multi-agent coordination tasks. Just as behavioral diversity strengthens agent performance, environmental diversity enhances the agent's generalization capability in MAS \cite{mckee2022quantifying}.
\begingroup
\footnotesize
\begin{longtable}{|L{0.125\linewidth}|L{0.15\linewidth}|L{0.65\linewidth}|}
    \caption{Ten organizational structures of multi-agent systems. Each organizational structure is distinguished by the following characteristics: information flow, relationships among agents, and goals of the organization.} \label{tab:mas_org} \\
    \hline
    \textbf{Organizational structure} & \textbf{Characteristics} & \textbf{Description} \\
    \hline
    \textbf{Flat} & Information Flow & Agents have access to information, reducing reliance on a central authority for dissemination. Communication among agents tends to adopt a peer-to-peer or distributed approach, facilitating sharing and collaboration across the network. \cite{horling2004survey,argente2007supporting,dorri2018multi}
    \\\cline{2-3}
    & Relationships among agents & Agents maintain egalitarian relationships. Interactions among these agents involve cooperation, negotiation, and collaboration.  \cite{horling2004survey,dorri2018multi} \\\cline{2-3}
    & Goals of the organization & The organization may prioritize flexibility and responsiveness to changes in the environment. \cite{horling2004survey,argente2007supporting,dorri2018multi} \\
    \hline
    \textbf{Hierarchies} & Information Flow & Information flows top-down from a central authority, with formalized communication channels and well-defined reporting relationships. Lower-level agents may have limited access to information beyond their immediate responsibilities, relying on superiors for guidance and updates. \cite{horling2004survey} \\\cline{2-3}
    & Relationships among agents & Agents' relationships have lines of authority and subordination, with each agent having roles and responsibilities. Interactions are governed by rules. For instance, superiors make decisions, and subordinates follow directives. This dynamic is directive, with superiors guiding and instructing while subordinates execute tasks. \cite{horling2004survey} \\\cline{2-3}
    & Goals of the organization & The organization emphasizes efficiency, control, and standardization for consistency across levels. Decision-making authority is centralized at the top, with lower-level agents primarily responsible for implementation rather than significant involvement in decision-making. \cite{horling2004survey} \\
    \hline
    \textbf{Holarchies} \cite{koestler1968ghost} & Information Flow & The flow includes top-down guidance and bottom-up feedback, enabling adaptability and self-organization within each holon. While overarching goals may exist, lateral communication and feedback loops occur between holons at the same level and across different levels. Agents within each holon share information to achieve collective goals and exchange information with holons at other levels. \cite{horling2004survey,rodriguez2006holonic,serugendo2011self} \\\cline{2-3}
    & Relationships among agents & Agents interact vertically (with higher and lower levels) and horizontally (with peers). There's a balance between autonomy and integration, allowing agents to self-organize within holons while aligning actions with larger goals. Functional hierarchies may form within holons for coordination, alongside cross-functional collaborations across holons. \cite{horling2004survey,rodriguez2006analysis,serugendo2011self} \\\cline{2-3}
    & Goals of the organization & Organizational goals are emergent and dynamic, shaped by interactions among holons at different levels. Each holon's goals and strategies reflect its context, capabilities, and relationships. The organization values adaptability and resilience, fostering continuous learning and evolution within the holarchy. \cite{horling2004survey,rodriguez2006analysis,serugendo2011self} \\
    \hline
    \textbf{Coalitions} & Information Flow & Communication occurs within and between coalitions to coordinate activities and decision-making internally and externally. \cite{shehory1998multi,van2004coalition,horling2004survey} \\\cline{2-3}
    & Relationships among agents & Agents collaborate closely, pooling resources to address shared goals. There is a balance between autonomy and alignment, with agents working toward collective goals while retaining some independence. \cite{ketchpel1994forming,zlotkin1994coalition,horling2004survey} \\\cline{2-3}
    & Goals of the organization & Agents focus on achieving specific objectives or addressing common goals. The organization reflects collective needs and priorities, driven by a commitment to advancing mutual interests. A coalition is ceased once organizational needs no longer exist. Hence, flexibility and adaptability are important within a coalition structure. \cite{ketchpel1994forming,zlotkin1994coalition,van2004coalition,horling2004survey} \\ 
    \hline
    \textbf{Teams} & Information Flow & The flow may be both vertical and horizontal, with communication occurring between team members, team leaders, and external entities. \cite{horling2004survey,coviello2012distributed} \\\cline{2-3}    
    & Relationships among agents & Members work closely, leveraging strengths to achieve shared objectives with trust and mutual respect. Clear roles and responsibilities ensure effective contributions to the team's efforts. \cite{horling2004survey,gaston2005agent,dunin2011teamwork} \\\cline{2-3}  
    & Goals of the organization & The team structure aims to establish authority for decision-making, fostering flexibility to adapt to changing circumstances, and enhancing collaboration among agents to work effectively towards shared objectives. \cite{horling2004survey,gaston2005agent} \\
    \hline
    \textbf{Congregations} & Information Flow & An agent only exchange information with other agents in the same congregation. \cite{brooks2000introduction,horling2004survey} \\\cline{2-3}
    & Relationships among agents & Agents in a congregation are cooperative. They establish trust and mutual understanding to facilitate teamwork and conflict resolution. \cite{brooks2000introduction,horling2004survey} \\\cline{2-3}
    & Goals of the organization & A congregation aims find suitable collaborators for a specific problem by grouping agents with similar characteristics. This coupling of agents helps maximize their utility or reliability. \cite{brooks2002congregating,brooks2003congregation,horling2004survey} \\
    \hline
    \textbf{Societies} & Information Flow & Agents exchange information within a society. Information flows through broadcasting or decentralized peer-to-peer networks. The complexity of this flow depends on the society's size, interaction density among agents, and the level of centralization in information dissemination. \cite{horling2004survey} \\\cline{2-3}
    & Relationships among agents & A society is an open system. Agents' interactions are formalized using contracts. They form social networks, hierarchies, alliances, or affiliations based on shared interests or organizational goals. \cite{carter2002reputation,horling2004survey} \\\cline{2-3}
    & Goals of the organization & The society structure aims to guide agents' behaviors and interactions through the collective objectives, shared values and social constraints, such as social laws, norms or conventions. \cite{carter2002reputation,horling2004survey} \\
    \hline
    \textbf{Federations} & Information Flow & Agents or sub-groups have local information but share relevant data to achieve common objectives. Information flows through agent hubs or peer-to-peer networks. A coordinator agent integrates information, resolves conflicts, establishes the information flow among federations, and ensures consistency in shared information. \cite{horling2004survey}
    \\\cline{2-3}  
    & Relationships among agents & Agents within a federation operate independently but collaborate on shared goals, often characterized by negotiation and cooperation. \cite{horling2004survey}
    \\\cline{2-3}
    & Goals of the organization & The federation structure aims to optimize system performance and achieve collective tasks which all agents work towards. Furthermore, while the structure supports flexibility and adaptability in goal setting and pursuit, it ensures local actions of individuals contribute to global objectives. \cite{horling2004survey} \\
    \hline
    \textbf{Markets} & Information Flow & Agents exchange price signals, bids, asks and other market information to facilitate market contracts. Information flows through peer-to-peer networks. Negotiation protocols and pricing mechanisms aim to coordinate agents' actions and allocate resources or tasks \cite{horling2004survey} \\\cline{2-3}
    & Relationships among agents & A market is an open system. Agents in a market have buyer-seller relationships. They seek trading partners and negotiate or compete with other agents to seek out the most cost-effective solutions. Their interactions are formalized using contracts. \cite{brooks2002congregating,horling2004survey} \\\cline{2-3}
    & Goals of the organization & A market creates a competitive environment that fosters cost-effective solutions for complex problems and maximizes agent-specific goals. The market structure enables the flexibility of the system in utilizing available resources to meet market demand. \cite{brooks2002congregating,horling2004survey} \\
    \hline
    \textbf{Matrix} & Information Flow & An agent reports to many managers and communicates with many peers. \cite{horling2004survey} \\\cline{2-3}
    & Relationships among agents &  Agents have dual reporting lines to both functional and project managers, fostering cross-functional relationships. Interactions are governed by formal rules. \cite{horling2004survey} \\\cline{2-3}
    & Goals of the organization &  The structure specify how multiple authorities influence the behaviours of agents or agent groups. This enhances organizational flexibility and responsiveness to changing conditions. \cite{horling2004survey} \\
    \hline
\end{longtable}
\endgroup

Once an organizational structure is established, achieving consensus, and mutual agreement on a common value or state becomes essential for enabling information exchange among agents. This, in turn, requires an agent to formulate its Belief-Desire-Intention (BDI) model for coordination when addressing complex problems \blue{\cite{michael1987intention,rao1995bdi,georgeff1999belief,jakobson2006situation,saleem2022situation}. \textit{Beliefs} represent the knowledge the agent has about the world, \textit{Desires} are the goals or objectives the agent wants to achieve, and \textit{Intentions} are the plans and actions the agent commits to in order to fulfill its desires. Once the BDI model is established in each agent, they begin communicating with one another. This introduces the dynamic state of communication for each agent. To achieve consensus among the agents, Amirkhani and Barshooi \cite{amirkhani2022consensus} outlined three categories of dynamic models: the single-integrator dynamic model, the double-integrator dynamic model, and the high-order dynamic model. Furthermore,} two primary challenges in agent communication encompass data flow issues, such as data continuity and communication delays, and data quality concerns, including outliers, and missing, or incomplete information \cite{qin2016recent}. To overcome these challenges, the literature has explored nine communication protocols, including (1) leader-follower consensus, (2) group/cluster consensus, (3) scaled consensus, (4) finite-time consensus, (5) bipartite consensus, (6) sampled-data consensus, (7) quantized consensus, (8) network-based consensus, and (9) the combination of consensus with tracking mechanisms \cite{qin2016recent,li2019survey,amirkhani2022consensus}. Additionally, optimizing the frequency of information transactions and computing resources per transaction is crucial to enhance communication flow for efficient coordination. To achieve this goal, existing trigger mechanisms controlling consensus can be divided into two groups: event-based consensus control and time-based consensus control \cite{qin2016recent,li2019survey}. The former triggers communication based on situations exceeding a pre-defined threshold or changes of state, while the latter triggers communication periodically based on a pre-defined time interval or selected data samples. \blue{Moreover, while reaching consensus among agents, noise and uncertainties may affect both system performance and agent-specific performance. To address this challenge and ensure data quality, consensus control strategies such as robust consensus control and adaptive consensus control have been investigated. These techniques can be further categorized into two groups: feedback control and feed-forward control \cite{amirkhani2022consensus}.}

\section{Context-Aware Systems} \label{sec:cas}
\begin{figure*}
   \includegraphics[width=\linewidth]{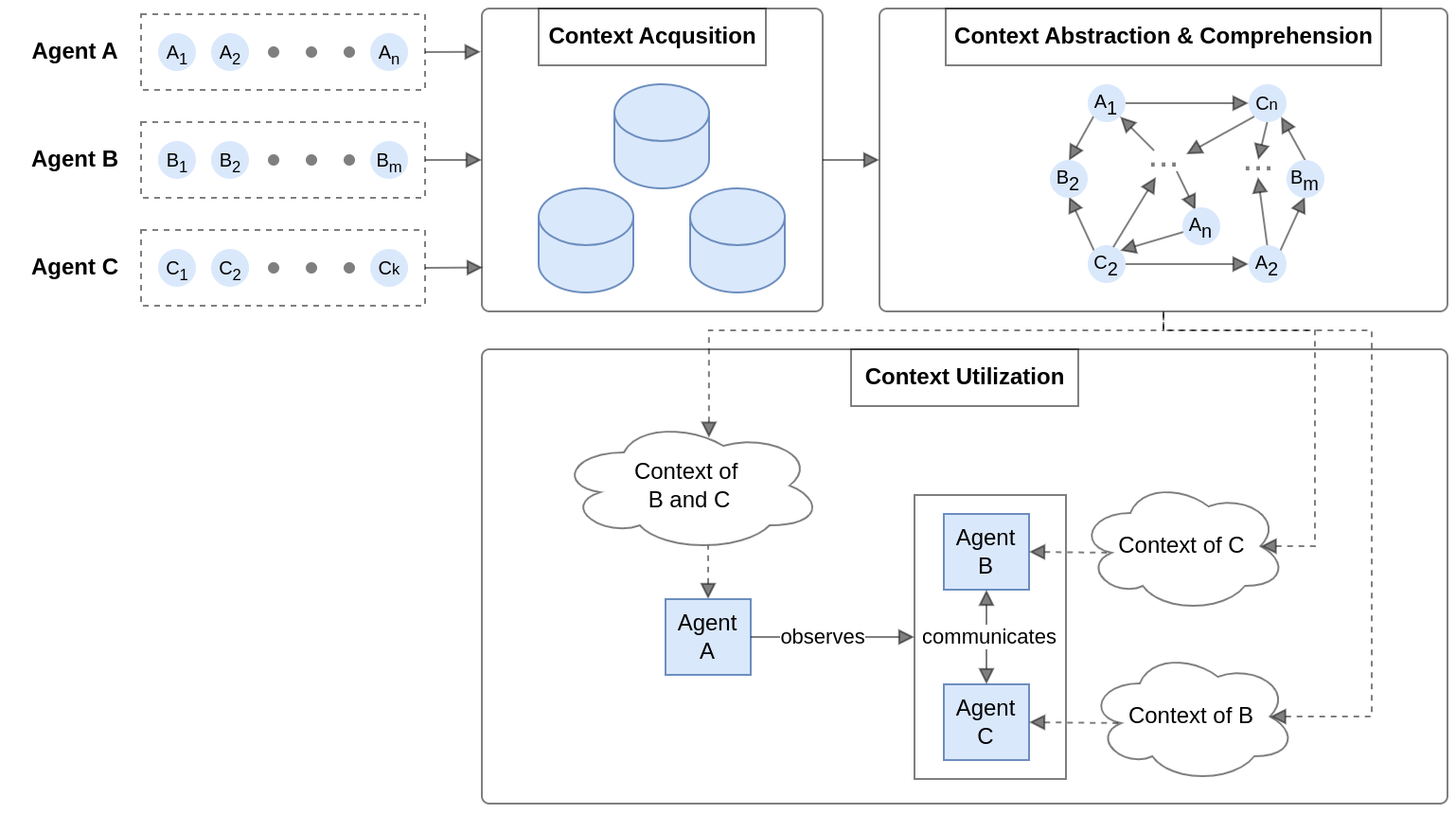}
   \caption{An general process of context-aware systems. A blue circle represents sensed information from an agent. During the context utilization stage, one possible scenario is where Agent A observes communication between Agents B and C. In this situation, Agent A observes and understands the communication between Agents B and C by retrieving their contexts and relationships. Similarly, Agents B and C understand each other's information by retrieving each other's contexts.}
   \label{fig:cas_overview}
\end{figure*}
Context encompasses various types of information, such as people, location, physical or virtual objects, events, time, and other data that can be employed to introduce different dimensions of a situation or conceptual information regarding specific circumstances \cite{abowd1999towards}. Additionally, five key properties characterize context: type, value, time when sensed, source where the information was gathered, and confidence in information accuracy \cite{baldauf2007survey}. Furthermore, context can be categorized into two groups: intrinsic context and extrinsic context. In the context of MAS, the former specifies the internal factors of an agent (e.g., goal, task, behavior, belief, knowledge, etc.), while the latter focuses on external factors, such as the environment, situation, social influence, and more. Existing context modeling techniques in the literature fall into six categories: (1) key-value models, (2) markup schema models, (3) graphical models, (4) object-oriented models, (5) logical-based models, and (6) ontology-based models \cite{strang2004context}. The selection of a context modeling technique \blue{depends on four factors, such as complexity, scalability, interoperability and reasoning assistance (see also Table \ref{tab:cm_compare}) \cite{strang2004context,bettini2010survey}}. \blue{For instance, key-value models are suitable for situations where simplicity is crucial, although they lack two following capabilities: capturing the relationships within context and reasoning. To address these challenges, other context modeling techniques are employed.}
It is worth noting that ontology-based context models are widely utilized due to their capability for semantic reasoning and representation of context relationships through knowledge graphs \cite{roussaki2006hybrid,baldauf2007survey,gu2020ontology,al2020ontology}. However, challenges associated with ontology-based context modeling include lack of generality \cite{gu2020ontology}, the difficulty in understanding ontology complexity and the substantial cost of maintaining ontologies \cite{hoareau2009modeling}. \blue{To address the challenge of generality, Gu \textit{et al.} \cite{gu2020ontology} proposed a context modeling technique that leverages the Web Ontology Language (OWL) to enhance semantic context representation. Additionally, Horrocks \textit{et al.} \cite{horrocks2004instance} introduced the InstanceStore system, which employs relational databases for semantic indexing and optimizes the performance of ontology-based reasoning}. While modeling or reasoning about context, it is crucial to consider three aspects: (1) the quality of context, including accuracy and completeness of information; (2) relationships between context elements; and (3) the flow of context, such as time-invariant context, time-variant context, and consistency in context switching \cite{strang2004context,hoareau2009modeling,bellavista2012survey}.
\begingroup
\footnotesize
\begin{longtable}{|C{0.15\linewidth}|C{0.11\linewidth}|C{0.11\linewidth}|C{0.15\linewidth}|C{0.11\linewidth}|L{0.22\linewidth}|}
    \caption{Characteristics and use cases of six categories of context modeling techniques. The rating scale includes Low, Medium and High. The best configuration features lower complexity, higher scalability, higher interoperability, and higher reasoning assistance. \cite{strang2004context,bettini2010survey}} \label{tab:cm_compare} \\
    \hline
    \textbf{Context Modeling} & \textbf{Complexity} & \textbf{Scalability} & \textbf{Interoperability} & \textbf{Reasoning Assistance} & \textbf{Use Cases} \\
    \hline
    \textbf{Key-value models} & Low & High & Low & None & Simple applications, basic context storage \\
    \hline
    \textbf{Markup schema models} & Medium & Medium & Medium & Medium & Hierarchical context, structured data \\
    \hline
    \textbf{Graphical models} & Medium & High & Medium & Medium & Complex relationships, networked contexts \\
    \hline
    \textbf{Object-oriented models} & Medium & High & Medium & Medium & Encapsulated context, object-oriented applications \\
    \hline
    \textbf{Logical-based models} & High & Medium & High & High & Context reasoning, rule-based systems \\
    \hline
    \textbf{Ontology-based models} & High & Medium & High & High & Semantic interoperability, advanced reasoning, and inference \\
    \hline
\end{longtable}
\endgroup

\begin{figure*}
   \includegraphics[width=\linewidth]{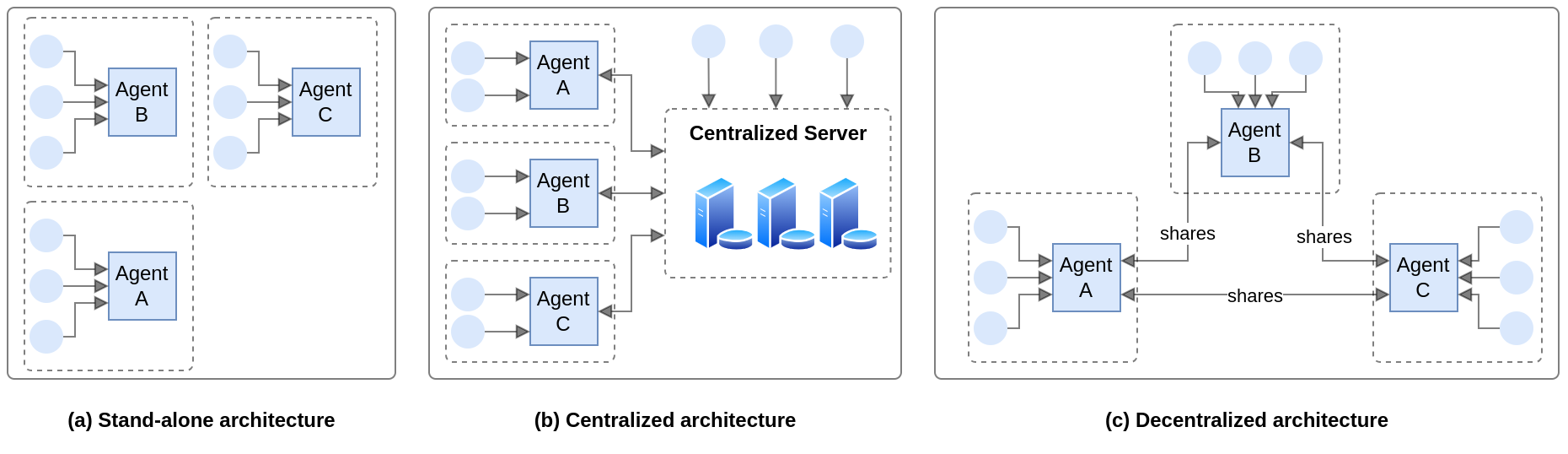}
   \caption{Architectures of context-aware systems. A blue circle represents the information generated by diverse sensors.}
   \label{fig:cas_arch}
\end{figure*}

Context-Aware Systems (CAS) are autonomous systems that leverage context to dynamically adapt to situations and retrieve relevant information for tasks or problems. Depending on the level of agent autonomy and interactions among agents, there are two types of context awareness: (1) passive context awareness, where an agent constantly produces observations from the environment that can be utilized by other agents; and (2) active context awareness, where an agent continuously and autonomously acts based on its observations from the environment \cite{perera2013context}. As illustrated in Figure \ref{fig:cas_overview}, the general process of CAS consists of three stages: (1) context acquisition, (2) context abstraction and comprehension, and (3) context utilization \cite{lee2011survey,perera2013context}. Context acquisition involves the process of sensing and storing information. In the context of MAS, an agent can sense information from its knowledge, other agents, or the environment. It is noteworthy that an agent may sense multiple types of context, periodically coming from various sources \cite{baldauf2007survey,hoareau2009modeling,perera2013context}. Based on the architectural style, \blue{Figure \ref{fig:cas_arch} shows three categories of context-aware systems \cite{lee2011survey}:} 
\begin{itemize}
    \item \textbf{Stand-alone CAS:} \blue{An agent retrieves information from sources of context and store such contextual information independently.}
    \item \textbf{Centralized CAS:} \blue{Contextual information is collected and stored in the centralized server. Agents can exchange the information with the server.}
    \item \textbf{Decentralized CAS:} This architecture does not rely on a central server; instead, it comprises multiple interconnected \blue{agents, each functioning as a stand-alone CAS and exchanging the information with others in the network.}
\end{itemize}
Furthermore, it is crucial to recognize that contextual information obtained from agents constitutes raw data. As a result, this data requires pre-processing and encapsulation to enable the interpretation of its semantics for a specific task or problem, a procedure commonly referred to as context abstraction and comprehension. This process involves techniques that can be classified into two types: context modeling and context reasoning. Context reasoning techniques, in particular, provide agents with the capability to infer knowledge from imperfect context and uncertainty. These techniques can be further categorized into six groups: (1) supervised learning, (2) unsupervised learning, (3) rule-based algorithms, (4) fuzzy logic, (5) ontology-based reasoning, and (6) probabilistic reasoning \cite{bettini2010survey,perera2013context,pradeep2019mom}. During the context utilization stage, an agent employs contextual information through its triggering mechanisms \cite{schilit1994context,brown1997context,perera2013context,alegre2016engineering}. Additionally, this information is distributed to other agents, guiding their decisions in solving complex tasks.

\section{Context-aware Multi-agent Systems} \label{sec:ca_mas}
Addressing time-varying uncertainties, such as the dynamics of an environment and the non-linearity of autonomous agents, poses a significant challenge in MAS. Overcoming this challenge entails agents being aware of changes in the environment or behavioral shifts of other agents within the system and consistently updating their beliefs based on observations. Achieving this goal requires the integration of CAS and MAS, resulting in what is termed \textit{Context-Aware Multi-Agent Systems (CA-MAS)}.
\begingroup
\footnotesize
\begin{longtable}{|L{0.15\linewidth}|L{0.2\linewidth}|L{0.57\linewidth}|}
    \caption{Existing CA-MAS approaches across eight application domains, including autonomous navigation, ambient intelligence, supply chain management, internet of things, disaster relief management, energy efficiency and sustainability, digital assistance, and education. Four characteristics of CA-MAS include organizational structure, consensus protocol, context modeling, and reasoning and planning approach.} \label{tab:ca_mas_app} \\
    \hline
    \textbf{Application Domains} & \textbf{Characteristics} & \textbf{References} \\
    \hline
    Autonomous navigation & Organizational Structure & Flat \cite{qi2018intent,everett2018motion,chen2020delay,tao2020dynamic,chen2021midas,xie2021congestion,lee2022density,fan2023switching,mahajan2023intent,wu2023intent} \\\cline{2-3}
    & Consensus Protocol & Sampled-data consensus \cite{qi2018intent,everett2018motion,chen2020delay,tao2020dynamic,chen2021midas,xie2021congestion,lee2022density,fan2023switching,mahajan2023intent,wu2023intent}  \\\cline{2-3}
    & Context Modeling & Key-value models \cite{qi2018intent,everett2018motion,chen2020delay,tao2020dynamic,chen2021midas,xie2021congestion,lee2022density,fan2023switching,mahajan2023intent,wu2023intent} \\\cline{2-3}
    & Reasoning and Planning & Goal-oriented algorithms \cite{qi2018intent,everett2018motion,chen2020delay,tao2020dynamic,chen2021midas,fan2023switching,mahajan2023intent,wu2023intent,xie2021congestion,lee2022density} \\
    \hline
    Ambient intelligence & Organizational Structure & Teams \cite{olaru2013context}, coalitions \cite{haiouni2019context} \\\cline{2-3}
    & Consensus Protocol & Group/cluster consensus \cite{olaru2013context,haiouni2019context} \\\cline{2-3}
    & Context Modeling & Ontology-based models \cite{olaru2013context}, key-value models \cite{haiouni2019context} \\\cline{2-3}
    & Reasoning and Planning & Graph-based algorithms \cite{olaru2013context}, rule-based algorithms \cite{haiouni2019context} \\
    \hline
    Supply chain management & Organizational Structure & Markets \cite{kwon2004applying}, holarchies \cite{fu2015adaptive} \\\cline{2-3}
    & Consensus Protocol & Sampled-data consensus \cite{kwon2004applying}, group/cluster consensus \cite{fu2015adaptive} \\\cline{2-3}
    & Context Modeling & Key-value models \cite{kwon2004applying}, ontology-based models \cite{fu2015adaptive} \\\cline{2-3}
    & Reasoning and Planning & Case-based techniques \cite{kwon2004applying}, hyrbid (rule-based algorithms and case-based techniques) \cite{fu2015adaptive} \\
    \hline
    Internet of things & Organizational Structure &  Teams \cite{twardowski2015iot}, flat \cite{huang2022intent}, coalitions \cite{yousaf2023context} \\\cline{2-3}
    & Consensus Protocol & Hybrid (group/cluster consensus and sampled-data consensus) \cite{twardowski2015iot}, sampled-data consensus \cite{huang2022intent}, group/cluster consensus \cite{yousaf2023context} \\\cline{2-3}
    & Context Modeling & Key-value models \cite{twardowski2015iot}, ontology-based models \cite{huang2022intent,yousaf2023context} \\\cline{2-3}
    & Reasoning and Planning & Goal-oriented algorithms \cite{twardowski2015iot, huang2022intent}, rule-based algorithms \cite{yousaf2023context} \\
    \hline
    Disaster relief management & Organizational Structure & Hybrid (teams and holarchies) \cite{jakobson2006situation,nadi2017adaptive,yusuf2022formalizing} \\\cline{2-3}
    & Consensus Protocol & Leader-follower consensus \cite{jakobson2006situation}, hybrid (leader-follower consensus and sampled-data consensus) \cite{nadi2017adaptive}, group/cluster consensus \cite{yusuf2022formalizing} \\\cline{2-3}
    & Context Modeling & Object-oriented models \cite{jakobson2006situation}, key-value models \cite{nadi2017adaptive}, logic-based models \cite{yusuf2022formalizing} \\\cline{2-3}
    & Reasoning and Planning & Hybrid (rule-based algorithms and case-based techniques) \cite{jakobson2006situation}, goal-oriented algorithms \cite{nadi2017adaptive}, graph-based algorithms \cite{yusuf2022formalizing} \\
    \hline
    Energy efficiency and sustainability & Organizational Structure & Holarchies \cite{yan2018energy}, teams \cite{jelen2022multi}, hierarchies \cite{riabchuk2022utility} \\\cline{2-3}
    & Consensus Protocol & Leader-follower consensus \cite{yan2018energy}, group/cluster consensus \cite{jelen2022multi}, hybrid (leader-follower consensus and sampled-based consensus) \cite{riabchuk2022utility} \\\cline{2-3}
    & Context Modeling & Key-value models \cite{yan2018energy,riabchuk2022utility}, object-oriented models \cite{jelen2022multi} \\\cline{2-3}
    & Reasoning and Planning & Goal-oriented algorithms \cite{yan2018energy, riabchuk2022utility}, rule-based algorithms \cite{jelen2022multi} \\
    \hline
    Digital assistance & Organizational Structure &   Flat \cite{fuentes2006reputation}, teams \cite{castellano2014multi} \\\cline{2-3}
    & Consensus Protocol & Sampled-data consensus \cite{fuentes2006reputation}, group/cluster consensus \cite{castellano2014multi} \\\cline{2-3}
    & Context Modeling & Ontology-based models \cite{fuentes2006reputation,castellano2014multi} \\\cline{2-3}
    & Reasoning and Planning & Graph-based algorithms \cite{fuentes2006reputation}, fuzzy rule-based algorithms \cite{castellano2014multi} \\
    \hline
    Education & Organizational Structure &  Hierarchies \cite{vladoiu2010learning} \\\cline{2-3}
    & Consensus Protocol & Group/cluster consensus \cite{vladoiu2010learning} \\\cline{2-3}
    & Context Modeling & Ontology-based models \cite{vladoiu2010learning} \\\cline{2-3}
    & Reasoning and Planning & Case-based techniques \cite{vladoiu2010learning} \\
    \hline
\end{longtable}
\endgroup

\subsection{Awareness in MAS}
Agent-based awareness techniques can be categorized into three groups \cite{mostafa2013formulating}: (1) Active perception \cite{bajcsy1988active,bajcsy2018revisiting,veiga2023reactive}, \blue{where an agent chooses what, when, where, and how to perceive, and can decide whether or not act based on its observations in an environment}; (2) situation awareness \cite{franklin1996agent,baader2009novel}, which involves reasoning about the current event and predicting future events as additional capabilities for agents to perform an optimal set of actions in a particular situation; and (3) context awareness, where context is utilized as the semantics of an event or a situation to enhance the agent's comprehension (see also Section \ref{sec:cas}). 
\blue{In this survey, we focus on the context-aware multi-agent systems.}

In MAS, \blue{an agent is considered context-aware when its behavior and actions adapt to the context it senses. Additionally,} an agent perceives both intrinsic and extrinsic context. The intrinsic context involves an agent's goals, roles, historical data (e.g., knowledge, previous actions), intent, and observations. Extrinsic context is categorized into user-specific context (e.g., location, preferences, calendar, weather, behavior from the user), agent-specific context (knowledge or intent sharing among agents, observations about roles or behavior of other agents), and system-specific context (system requirements, policies, organizational structures, and communication protocols followed by agents for optimal problem-solving in specific situations). \blue{Additionally, Picard \textit{et al.} \cite{picard2009reorganisation} emphasized the impact of agents' organizational awareness on their adaptation, flexibility, and coordination. Table \ref{tab:ca_mas_app} illustrates that various organizational structures, such as flat, teams, coalitions, holarchies, markets, and hierarchies (see also Section \ref{sec:mas}), are primarily utilized to arrange CA-MAS across diverse application domains. Within these structures, the consensus protocols among agents include sampled-data consensus, group/cluster consensus, and leader-follow consensus.}
Agent-specific context involves knowledge or intent sharing among agents and an agent's observations about roles or behavior of other agents. It is noteworthy that agents in CA-MAS can be either communicating agents or non-communicating agents \cite{chen2017decentralized,everett2018motion}. A communicating agent exchanges information or intent with other agents in the system via the communication protocol. On the other hand, a non-communicating agent does not exchange information with other agents \blue{via a communication protocol}; however, its behavior can be sensed or predicted based on its actions. For instance, in a competitive MAS setting, an agent performs actions based on observations and predictions about the actions of opponents \cite{nezamoddini2022survey}. Another example is in autonomous driving agents, where, to avoid collisions on the road, agents must act based on observations \cite{chen2017decentralized,everett2018motion,xie2021congestion,wu2023intent}.

\subsection{The General Process of CA-MAS} \label{sec:process_ca_mas}
The general process of CA-MAS comprises five phases: Sense-Learn-Reason-Predict-Act \cite{bruce2004better,dobson2006survey,rudenko2020human,xie2021congestion}. Details of each phase are specified below. 

\begin{figure*}
   \includegraphics[width=\linewidth]{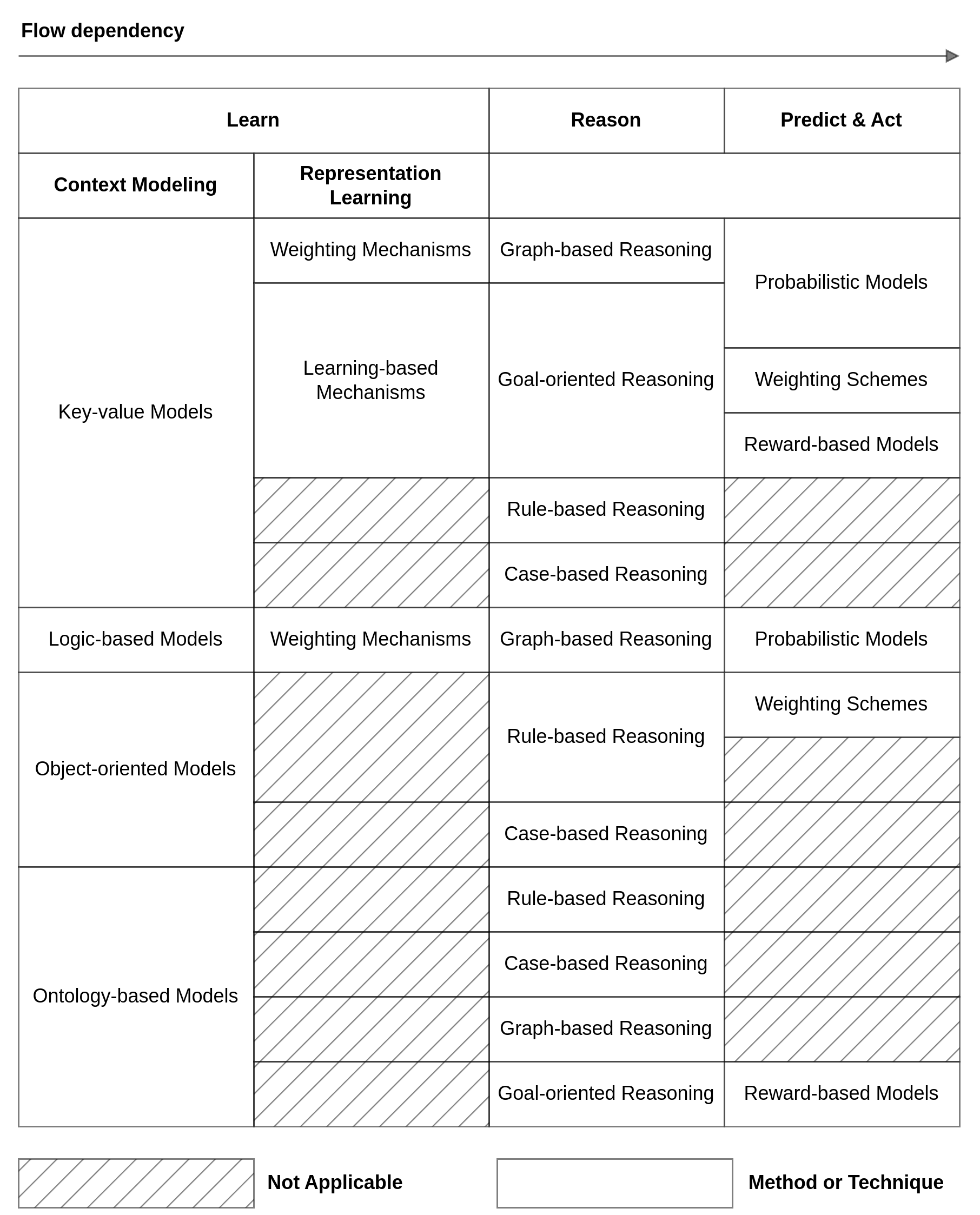}
   \caption{The flow dependency of techniques in Learn, Reason, Predict and Act. Note that the reading order is from left to right.}
   \label{fig:camas-process-group}
\end{figure*}

\subsubsection{Sense} \label{sec:camas_sense}
An agent in MAS gathers contextual information from its sensors, detects context patterns, and handles temporal changes of context \blue{(see also Figure \ref{fig:cas_overview})}. Context can be sensed from a stand-alone node, a centralized server, or through communication and interaction among agents \blue{(see also Figure \ref{fig:cas_arch})}. \blue{As the context is sensed from multiple sources, relationships between contexts may exists. These relationships are, then, represented as the context graph denoted as $G = (V, E)$ where $V$ is a set of context and the edge $E$ is a subset of $V \times V$. Note that there are two types of $G$: (1) a directed graph where a directed edge is represented as an ordered pair of context $c_{i}$ and $c_{j}$ and denoted by ($c_{i}$, $c_{j}$) or $c_{i} \rightarrow c_{j}$; and (2) an undirected graph where an edge between $c_{i}$ and $c_{j}$ is denoted by ($c_{i}$, $c_{j}$) $\in E$ and interprets the relationship between those context. Additionally, a context pattern is defined as a subgraph of $G$ and denoted as $H = (V', E')$ where $V' \subseteq V$ and $E' \subseteq E$. This pattern supports an agent to make decisions in a specific situation. Furthermore, the constant change in the environment increases the number of contexts, followed by the number connections in the context graph as:
\begin{equation}
    \text{N}_{E} = \frac{\text{N}_{V} \times (\text{N}_{V} - 1)}{2}
\end{equation}
where N$_{E}$ is the number of connections, and N$_{V}$ is the number of contexts on the graph. This leads to a high complexity in the search space, even though an agent might only need to concentrate on a specific subset of the context to solve a task.}
To overcome the challenge of managing context changes in dynamic environments, Julien \textit{et al.} \cite{julien2004supporting} applied the Network Abstractions model \cite{roman2002network}. \blue{The model estimates the weighted distance, denoted as $d(c_{r}, c_{k})$, between the reference context ($c_{r} \in V$) and another context ($c_{k} \in V$) by calculating the shortest path between them as \cite{roman2002network}:
\begin{equation} \label{eq:nabs_1}
    d(c_{r}, c_{k}) = \min{(d(c_{r}, c_{k - 1}) + w(c_{k-1}, c_{k}))} 
\end{equation}
where $d(c_{r}, c_{k - 1})$ is the minimum weighted distance of all possible paths between $c_{r}$ and $c_{k-1}$, and $w(c_{k-1}, c_{k})$ is the cost of the edge between $c_{k-1}$ and $c_{k}$. The cost function of the model is, then, estimated by determining both $d(c_{r}, c_{k})$ and the number of hops between $c_{r}$ and $c_{k}$ as \cite{roman2002network}:
\begin{equation} \label{eq:nabs_2}
    u = (D_{\max}, H_{\max}, V, h)
\end{equation}
where $D_{\max}$ represents the maximum distance between $c_{r}$ and $c_{k}$, $H_{\max}$ indicates the maximum number of consecutive hops during which $D_{\max}$ has not changed, $V$ is the distance vector capturing the displacement and direction between $c_{r}$ and $c_{k}$, and $h \leq H_{\max}$ denotes the count of consecutive hops for which $D_{\max}$ has remained unchanged. When an agent navigates in the dynamic environment, $c_{r}$ is shifted. This results in the update of Equations \ref{eq:nabs_1} and \ref{eq:nabs_2}. Hence, the model helps the agent handle context changes from the dynamics of the environment.
} 

\subsubsection{Learn} \label{sec:camas_learn}
An agent learns representations of information along with the sensed context for a specific task. \blue{This procedure involves two stages: context modeling (refer to Section \ref{sec:cas}) and representation learning, which includes weighting mechanisms and learning-based approaches.}

\blue{Key-value models, object-oriented models, logic-based models, and ontology-based models have been widely used in existing CA-MAS across application domains (see also Table \ref{tab:ca_mas_app}). Due to the simplicity and scalability of the key-value model, it has been primarily utilized in autonomous navigation to manage various types of contextual information: (1) the state representing the environment's configuration, denoted as $s \in \mathcal{S}$; (2) the action of the agent, denoted as $a \in \mathcal{A}$; and (3) the observation sensed by the agent, denoted as $o \in \mathcal{O}$ \cite{qi2018intent,everett2018motion,chen2020delay,tao2020dynamic,chen2021midas,xie2021congestion,fan2023switching,mahajan2023intent,wu2023intent}. Similarly, Nadi and Edrisi \cite{nadi2017adaptive} employed key-value pairs to model the states ($s$) and actions ($a$) of agents within the field of disaster relief management. Furthermore, the simplicity of key-value pairs facilitates the vectorization of context, represented by $\vec{c} = (c_{1}, c_{2}, \cdots, c_{n})$, where $\vec{c}$ is a vector encompassing contextual information. This approach has been applied in various domains such as autonomous navigation \cite{lee2022density}, supply chain management \cite{kwon2004applying}, the Internet of Things \cite{twardowski2015iot}, and energy efficiency \& sustainability \cite{yan2018energy,riabchuk2022utility}. Furthermore, Wan and Alagar \cite{wan2008context} utilized key-value models to capture both internal and external contexts of agents, associating these contexts with trust attributes, such as safety, security, reliability, and availability, to support the modeling of trust among agents forming a coalition or team. Moreover, Haiouni et al. \cite{haiouni2019context} applied key-value pairs to model the world state in a set-theoretic planning domain \cite{ghallab2004automated}. Apart from key-value models, CA-MAS which employs an object-oriented model to represent contexts such as weather forecast data and points of interest data was proposed by \cite{jelen2022multi} to optimize planning costs and resource utilization for electric vehicle fleet routing in urban operations. Furthermore, Yusuf and Baber \cite{yusuf2022formalizing} utilized logic-based models to capture the relationships between various contextual information, including types of fuel, fuel conditions, the existence of fire, and location within the domain of disaster relief management. When the semantics of contextual information are required to model complex situations or problems, ontology-based models are employed \cite{fuentes2006reputation,vladoiu2010learning,olaru2013context,castellano2014multi,fu2015adaptive,huang2022intent,yousaf2023context}.
}

CA-MAS with weighting mechanisms \blue{are} categorized into two groups: statistical approaches \cite{kwon2004applying,twardowski2015iot,yan2018energy,lee2022density,riabchuk2022utility} and graph-based approaches \cite{julien2004supporting,yusuf2022formalizing}. \blue{Furthermore}, the weighting mechanism encompasses three components: (1) vector representation, where context values are transformed into a machine-readable format and stored in a vector; (2) weight estimation, where weights are measured by aggregating between context vectors and other vectors representing other entities in the system; and (3) weight utilization, where estimated weights are used for goal-oriented optimization. \blue{Kwon and Sadeh} \cite{kwon2004applying} proposed an approach that estimates the weight of an agent by aggregating between the vector of contextual information and the utility vector of that agent for automating negotiation tasks between buyers and sellers \blue{and maximize their payoff}. Furthermore, \blue{Lee and Kabir} \cite{lee2022density} proposed the density-aware MAS that utilizes agents' density distribution as context vectors to optimize the spatial exploration of a team of robots. Moreover, domain-specific contextual information, such as utility usage \cite{yan2018energy,riabchuk2022utility} and mobile usage \cite{twardowski2015iot}, gathered from agents, is stored in a vector for weight estimation in recommendation systems. 
Aside from statistical approaches, Yusuf and Baber \cite{yusuf2022formalizing} \blue{represented each type of contextual information collected by unmanned aerial vehicles (UAVs) as a node within a Bayesian Belief Network (BBN) and calculated their joint probability as \cite{pearl1988probabilistic}:
\begin{equation}
    p(c_{1}, c_{2}, \cdots, c_{n}) = \prod_{i = 1}^{n}{p(c_{i} | B(c_{i}))}
\end{equation}
where $c_i$ denotes a context node in the directed acyclic graph $G$ (see also Section \ref{sec:camas_sense}), and $B(c_i)$ represents the set of parent nodes of $c_{i}$. Specifically, the joint probability of three contexts ($c_{1}$, $c_{2}$, and $c_{3}$), where $c_{1}$ and $c_{2}$ are parents of $c_{3}$, is determined as follows:
\begin{equation}
\begin{aligned}
    \quad & p(c_{1}, c_{2}, c_{3}) = p(c_{1}) \times p(c_{2}) \times p(c_{3} | c_{1}, c_{2}) \\
    \text{subject to} \quad & p(c_{3} | c_{1}, c_{2}) = \frac{p(c_{1}, c_{2} | c_{3}) \times p(c_{3})}{p(c_{1}, c_{2})} \\
\end{aligned}
\end{equation}
The probability $p\left(c_i | B(c_i)\right)$ represents the likelihood of a context node given its parent nodes.
Yusuf and Baber \cite{yusuf2022formalizing} assigned probabilities to each contextual information based on its likelihood of occurrence.
}

\begin{figure*}
   \includegraphics[width=\linewidth]{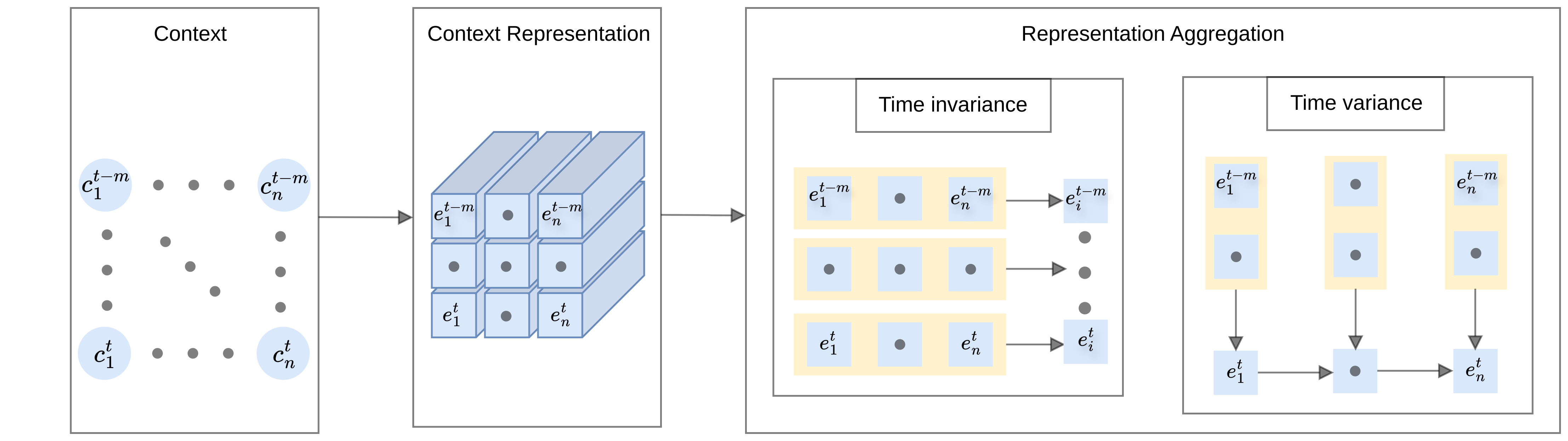}
   \caption{The general process of learning-based approach in existing CA-MAS. $n$ is a number of agents, $t$ is the current time, and $t - m$ indicates $m$ steps before $t$. Note that there are two main settings of aggregation: (1) time invariance illustrating the aggregation process is not influenced by time; and (2) time variance showing historical data of each agent is considered.}
   \label{fig:ca-mas-learn}
\end{figure*}
As the agent's observation space expands, the volume of context vectors and their dimensions increase, leading to the high-dimensional dilemma, also known as the curse of dimensionality \cite{bellman1961adaptive,donoho2000high,verleysen2005curse,mnih2015human,menda2018deep}.
Furthermore, due to uncertainty and ambiguity, integrating the weighting scheme with goal-oriented reasoning frameworks (see also Section \ref{sec:reason}) becomes challenging. Overcoming these challenges necessitates the use of \blue{learning-based} techniques for two reasons. First, contextual information can be represented in low-dimensional vector space. Second, learning parameters can be optimized according to a given goal or policy. \blue{Figure \ref{fig:ca-mas-learn} illustrates the overall process of representing contextual information in MAS using learning-based methods. The vector representation of time-invariant context, represented as $e$, can be determined as follows:
\begin{equation}
    e = \phi(c, W_{c})
\end{equation}
where  $\phi$ represents the embedding function with a non-linear approximation, and $W_{c}$ denotes the embedding weights. The selection of the embedding function is influenced by the modality of the context \cite{pouyanfar2018survey,gao2020survey}. Since an agent can perceive various contexts from other agents, dependencies between contexts may exist and can be calculated as follows:
\begin{equation}
    e_{i} = \phi\left(\bigcup_{j=0}^{N}{e_{j}}, W_{l}\right)
\end{equation}
where $\bigcup$ indicates the combination of a set of embedding vectors, $N$ is the number of contexts, $i$ and $j$ are the indices of $c_{i}$ and $c_{j}$ respectively, and $W_{l}$ represents the embedding weights for aggregation. Moreover, the vector representation of time-variant context can be estimated as follows:
\begin{equation}
    e^{t}_{i} = \phi\left(\bigcup_{j=0}^{N}{e^{t - 1}_{j}}, W_{k}\right)
\end{equation}
where $t$ and $t - 1$ represents the current state and the past state of the environment, respectively. An agent's observations from the traffic environment are represented using a Multi-Layer Perceptron (MLP) as follows: $e = \text{MLP}(o)$. Similarly, the CA-MAS model proposed by Fan et al. \cite{fan2023switching} employed an MLP with an extraction layer \cite{he2016deep} to encode an agent's observations of obstacles in the multi-agent particle environment \cite{lowe2017multi}. This method also utilized a 1D convolutional network to aggregate embedding vectors of an agent's observations and actions to select an optimal action. Additionally, in the CA-MAS model proposed by Qi and Zhu \cite{qi2018intent}, YOLO \cite{redmon2016you} was used to detect visual objects in the simulated V-REP environment \cite{rohmer2013v}, aiding the agent's path planning activities. Everett et al. \cite{everett2018motion} formulated the agent's state as a sequence of information comprising two types: observable states and unobservable states, denoted by $s^{o}$ and $s^{h}$, respectively. Specifically, $s^{o}$ includes the agent's position, velocity, and radius, while $s^{h}$ includes the goal position, preferred speed, and orientation. Everett et al. \cite{everett2018motion} further aggregated the agent's state with those of other agents, represented using Long Short-Term Memory (LSTM) \cite{hochreiter1997long}, as follows:
\begin{equation}
e = \phi\left(\bigcup(s, \text{LSTM}(\tilde{s^{o}{1}}, \cdots, \tilde{s^{o}{n}})), W_{s}\right)
\end{equation}
where $e$ is the aggregated state embedding of the agent, $W_{s}$ indicates the embedding weights of the aggregation, $\tilde{s}$ represents the other agent's state, and $n$ is a number of agents. LSTM was also applied in the CA-MAS model proposed by Tao et al. \cite{tao2020dynamic} to represent both the agent's observations and those of other agents. Beyond features across agents at a specific time, this approach also incorporates historical information of successive events. These representations were then aggregated with the semantic representation of the latest observed motions, enabling the agent to learn both the spatial dependencies and temporal coherence of moving agents. 
}
Pedestrian trajectories are also crucial for estimating uncertainty in autonomous driving. To achieve this goal, \blue{Katyal \textit{et al.}} \cite{katyal2020intent} proposed a CA-MAS where the integration of pedestrian trajectories and agents' observations was represented via the Encoder-Decoder architecture consisting of stacked LSTM. 
\blue{
Moreover, to help the agent learn the correlation between the control actions of other agents, Chen and Chaudhari \cite{chen2021midas} introduced a CA-MAS in autonomous navigation. They used an MLP to represent driver types and an Induced Set Attention Block (ISAB) from the Set Transformer \cite{lee2019set} to encode the control actions of autonomous vehicles along with their correlations.
}
Aside from using pedestrian trajectories as contextual cues, \blue{Xie \textit{et al.}} \cite{xie2021congestion} utilized congestion patterns as context, which were represented by Graph Convolution Network integrated with Variational Auto-Encoder (GCN-VAE) \cite{Kingma2013AutoEncodingVB,kipf2016semi} for trajectory prediction.
\blue{
Furthermore, Wu \textit{et al.} \cite{wu2023intent} proposed CA-MAS in the heterogeneous
traffic where the agent needs to understand other agents' driving actions for making instant response. In this system, Gated Recurrent Unit (GRU) \cite{chung2014empirical} was applied to represent agents' historical observations and driving actions, denoted by $o_{i}^{t - 1}$ and $\beta$, respectively. Additionally, Graph Attention Network (GAT) \cite{velivckovic2017graph} was employed to extract the temporal relationship between $o_{i}^{t}$ and $\beta$.
}

\subsubsection{Reason} \label{sec:reason} \label{sec:camas_reason}
Agents analyze information or formulate a set of plans based on the sensed context to achieve their objectives. \blue{As shown in Table \ref{tab:ca_mas_app}}, various reasoning models, including \blue{rule-based reasoning, case-based reasoning, graph-based reasoning, and goal-oriented reasoning have been employed.}

Rule-based reasoning involves a predefined set of rules governing how agents respond to specific conditions or events. In the context of CA-MAS, \blue{Wan and Alagar} \cite{wan2008context} incorporated a set of rules to model trust, while \blue{Ferrando and Onaindi} \cite{ferrando2013context} utilized defeasible rules to formulate arguments supporting the planning process. To enhance the reasoning capabilities of agents in complex situations, \blue{Haiouni and Maamri} \cite{haiouni2019context} proposed the Hierarchical Colored Petri Net, an extension of the Petri Net \cite{murata1989petri}, which applies agent hierarchies to the planning process in CA-MAS. In the domain of urban planning, \blue{Jelen \textit{et al.}} \cite{jelen2022multi} implemented a rule-based routing algorithm using contextual information to identify urban parking lots or charging stations for electric vehicles. Meanwhile, \blue{Yousaf \textit{et al.}} \cite{yousaf2023context} enhanced agent reasoning capabilities by applying rules to identified matching patterns in the context ontology. \blue{Furthermore, agents utilize their world model and intrinsic motivation (see also Section \ref{sec:mas}) to formulate plans supporting their goals. To represent the mental states of agents, the BDI model is applied. Jakobson \textit{et al.} \cite{jakobson2006situation} extended the traditional BDI model by adding the Plan module, which consists of context ontology and enhances agents' recognition of dynamic situations.}
An extension of rule-based reasoning is fuzzy logic, where agents can represent and process information with a certainty degree for a particular situation based on semantic rules. In this regard, \blue{Castellano \textit{et al.}} \cite{castellano2014multi} employed a neuro-fuzzy network, representing fuzzy rules as a neural network architecture to map contextual information to specific situations. 

Case-Based Reasoning (CBR) empowers agents to retrieve, reuse, revise, and retain cases that represent their past experiences. This aims to support agents to adapt and make decisions in similar situations \cite{watson1994case}. \blue{Given the agent's contextual information, retrieving a set of all relevant cases can be formulated as:
\begin{equation}
    f(c) = \{v \in V \mid g(c, v) \geq \alpha\}
\end{equation}
where $f$ is the retrieval function that identifies relevant cases based on the current context $c$, $v$ is a case in a set of all possible cases $V$, $g$ is the similarity function that measures how closely a case $v$ matches the current context $c$, and $\alpha$ is the similarity threshold. 
} In CA-MAS proposed by \blue{Kwon and Sadeh} \cite{kwon2004applying}, CBR is applied to retrieve similar cases based on contextual information during negotiation tasks. In the domain of multi-dimensional learning, \blue{Vladoiu and Constantinescu} \cite{vladoiu2010learning} employed CBR with agents' ontology to retrieve learning scenarios tailored to the user's interests. Furthermore, \blue{Fu and Fu} \cite{fu2015adaptive} equipped agents with both rule-based reasoning and CBR, integrated on top of their context ontology. This integration allows agents to derive high-level business context from low-level contextual information for collaborative cost management in the supply chain. 

Graph-based reasoning endows agents with the capability to analyze complex interactions, dependencies, and patterns within graph structures. CA-MAS in \cite{fuentes2006reputation} personalized recommendations for users by matching their contextual information with patterns identified within agents' context ontology. Agents in both \cite{beydoun2009security} and \cite{feyzi2020model} utilized meta-models, encompassing agents' contextual information, relationships among such information, and inter-agent relationships, for reasoning in specific situations. In another approach, \blue{Olaru \textit{et al.}} \cite{olaru2013context} introduced the Tri-Graph, a union of graphs representing restricted environments, contextual information, and agent topology. This graph captures context patterns for \blue{matching, filtering and reasoning about information. Aside from pattern matching algorithms, a graphical context model with probabilistic estimation such as Bayesian Belief Networks (BBN) have been applied to support agents in reasoning about fire spread in the field of disaster relief management \cite{yusuf2022formalizing}.} 

In goal-oriented reasoning, agents prioritize and plan actions based on their goals. While other reasoning techniques that rely on pre-programmed conditions, patterns, or cases, agents with goal-oriented reasoning capability adapt to any situation to optimally achieve their goals. \blue{Additionally, every goal-oriented CA-MAS consists of an optimization process with a cost function. Riabchuk \textit{et al.} \cite{riabchuk2022utility} applied the mean squared error as the cost function that illustrate the accuracy of the typical energy consumption of operating a device in approximating the actual one:
\begin{equation}
    \mathcal{L}^{i} = \frac{1}{n}\sum_{d}\sum_{j}\frac{1}{k}{(y^{i}_{d,j} - \hat{y}^{i}_{d})^{\intercal}(y^{i}_{d,j} - \hat{y}^{i}_{d})}
\end{equation}
where $i$ is the index of the $i^{\text{th}}$ device, $n$ is the total number of uses of devices, $k$ is the usage duration in hours, $\hat{y}^{i}_{d}$ is a typical energy consumption of the $i^{\text{th}}$ device at usage date $d$, and $y^{i}_{d,j}$ is a true energy consumption of the $i^{\text{th}}$ device at usage date $d$ with index $j$. Similarly, Twardowski and Ryzko \cite{twardowski2015iot} employed the regularized square error to optimize the recommendation system of mobile usage based on user preferences as \cite{shi2014cars2}:
\begin{equation}
    \mathcal{L}(\theta) = \frac{1}{2}\sum_{(i, j, k) \in R}(y_{ijk} - \hat{y}_{ijk})^{2} + \Omega(\theta)
\end{equation}
where $i$, $j$, and $k$ represents the $i^{\text{th}}$ user, the $j^{\text{th}}$ item and the $k^{\text{th}}$ context, respectively, and $\Omega(\theta)$ is the regularization term which control magnitude of learning parameters $\theta$. Additionally, to find the best learning parameters, Stochastic Gradient Descent (SGD) \cite{robbins1951stochastic} was applied \cite{twardowski2015iot}. Furthermore, as a student in \cite{xie2021congestion}, the agent is required to reason about congestion patterns for collision avoidance taught by the teacher. Specifically, Gaussian Mixture Model (GMM) \cite{mclachlan1988mixture} was employed to estimate the representation of congestion patterns \cite{xie2021congestion}:
\begin{equation}
    Q(o) = \sum^{M_{Q}}_{i}{\lambda_{i}q_{i}(o)} 
\end{equation}
where $Q$ is the teacher's GMM, $q_{i}(o)$ is the $i^{\text{th}}$ mixture component in a Gaussian distribution, $\lambda_{i}$ is the mixture weight, and $M_{Q}$ is the hyperparameters of the total number of mixtures. Additionally, the student's GMM is denoted as $P(o) = \sum^{M_{P}}_{j}{\omega_{j}p_{j}(o)}$ where $M_{P}$ may differ from $M_{Q}$ \cite{xie2021congestion}. To support the reasoning procedure of a student, Xie \textit{et al.} \cite{xie2021congestion} aims to minimize the distance between $Q(o)$ and $P(o)$ as follows:
\begin{equation}
    \min_{\{p_{j}\}, \alpha, \beta}{\mathcal{L}} = \sum_{i, j}\alpha_{i, j}\mathbb{D}_{\text{KL}}(p_{j}(o) || q_{i}(o)) + \mathbb{D}_{\text{KL}}(\alpha || \beta)
\end{equation}
where $\mathcal{L}$ is L1 loss term, $\alpha$ and $\beta$ are obtained by decomposing the mixture weights $\omega_{j} = \sum_{i}^{M_{Q}}\alpha_{ij}$ and $\lambda_{i} = \sum_{j}^{M_{P}}\beta_{ij}$, respectively, and $\mathbb{D}_\text{DL}$ is the KL-divergence between two pattern distributions.
}
Moreover, reasoning with feedback, which is from both the environment and other agents, supports an agent to adapt to situations. To attain this, \blue{Reinforcement Learning} (RL) techniques are applied. In RL, an agent aims to maximize its rewards by selecting an optimal action based on a policy at a particular state \cite{sutton2018reinforcement}. 
\blue{
In conjunction to three properties in Section \ref{sec:camas_learn}, there are two additional properties in Multi-Agent Reinforcement Learning (MARL) determined by a fully observable Markov Decision Process (MDP) \cite{bellman1957markovian}. With $N$ agents, these properties include $\mathcal{S}$, $\mathcal{A} = \{ \mathcal{A}^{1}, \cdots, \mathcal{A}^{N} \}$,  $\mathcal{O} = \{ \mathcal{O}^{1}, \cdots, \mathcal{O}^{N} \}$, $\mathcal{P}$ denoting the transition probability between states, $\mathcal{R} = \{ \mathcal{R}^{1}, \cdots, \mathcal{R}^{N} \}$ representing a set of agent's rewards, and the discount factor $\gamma \in [0, 1)$. In Partially Observable Markov Decision Process (POMDP) \cite{cassandra1994acting,kaelbling1998planning} where the agent does not have full observability of the environment state, the agent maintains a belief about the state based on its observations and actions. Therefore, there are two additional properties, such as the observation probability and the belief state update function denoted by $\mathcal{Z}$ and $b'$, respectively. MARL approaches have been studied in the literature \cite{busoniu2008comprehensive,hernandez2019survey,zhang2021multi,gronauer2022multi}. In this survey, we focus on approaches that have been applied in CA-MAS. To support an agent to reason about unconventional situations and human requests in disaster relief management, Nadi and Edrisi \cite{nadi2017adaptive} utilized $n$-step Q-learning as \cite{watkins1992q}:
\begin{equation}
\begin{aligned}
    \quad & Q(s, a) = (1 - \alpha_{n})Q_{n - 1}(s, a) + \alpha_{n}[r(s, a) + \gamma\max_{a'}Q_{n - 1}(s', a')] \\
    \text{subject to} \quad & \alpha_{n} = \frac{1}{1 + N_{s, a}} \\
\end{aligned}
\end{equation}
where $s'$ and $a'$ are the future state and action, respectively, and $N_{s, a}$ represents the number of visited pairs of state and action until step $n$.  Moreover, understanding other agents' intents support the agent to construct its goals for maximizing its expected utility. To achieve this, Qi and Zhu \cite{qi2018intent} modeled the agent's goal with its utility function and parameterized that function by using linear function approximation as follows:
\begin{equation}
    g_{i} = \argmax_{g}u(g \mid b_{-i}; h, \theta)
\end{equation}
where $g_{i}$ represents the goal of the $i^{\text{th}}$ agent, $b_{-i}$ indicates the belief of intents of all other agents, $h$ is the observation history, and $\theta$ represents all intent combinations of agents. Note that $\theta = (\theta_{ik,jl}) \in \mathbb{R}^{m \times n \times m}$ represents the intrinsic value of an agent $i$. In this matrix, $m$ is the number of goals, $n$ is the number of agents, and $\theta_{ik,jl}$ denotes the utility for agent $i$ to pursue goal $g_{k}$ when agent $j$ is intended for goal $g_{l}$. Then, the utility function of the agent is estimated as \cite{qi2018intent}:
\begin{equation} \label{eq:agent_utility}
    u(g_{i,k} \mid b_{-i}; h, \theta) = \theta_{ik,ik} + \sum_{j \neq i}\sum_{l}\theta_{ik,jl}p(g_{jl} \mid h)
\end{equation}
where $p(g_{jl} \mid h)$ is the probability of the event of agent $j$ pursuing goal $g_{l}$ given the history. In RL setting, Equation \ref{eq:agent_utility} can be approximated as Q-value function as follows: $Q(h, g_{ik}) \approx  u(g_{i,k} \mid b_{-i}; h, \theta)$. Additionally, the semi-gradient State-Action-Reward-State-Action (SARSA) algorithm \cite{rummery1994line} was utilized to update $\theta$ as follows:
\begin{equation}
    \theta_{t+1} = \theta_{t} + \alpha\delta\nabla Q(h_{t}, g_{t}; \theta_{t})
\end{equation}
\begin{equation}
    \overline{r}_{t+1} = \overline{r}_{t} + \beta\delta
\end{equation}
where $\delta$ the estimation error of the average reward \cite{mahadevan1996average}, $\alpha$ and $\beta$ are learning rates. Furthermore, deep RL (DRL) techniques for goal-oriented reasoning were applied in existing CA-MAS \cite{everett2018motion,chen2020delay,chen2021midas,huang2022intent,fan2023switching,mahajan2023intent,wu2023intent} that employed learning-based techniques to represent high-dimensional sensory input. To support the agent to reason about sequences of human activities, the combination of Universal Value Function Approximators (UVFAs) \cite{schaul2015universal} and Policy Continuation with Highsight Inverse Dynamics (PCHID) \cite{sun2019policy} has been utilized in \cite{huang2022intent}. Specifically, UVFAs aim to approximate the reward associated with achieving specific goals $r_{g} : \mathcal{S} \times \mathcal{A} \rightarrow \mathbb{R}$. In addition, the technique formulates policies to get the optimal set of actions from state and goal pairs as follows: $\pi : \mathcal{S} \times \mathcal{G} \rightarrow \mathcal{A}$. PCHID aims to optimize the reasoning process after unifying $n$ steps of goals and policies \cite{huang2022intent}:
\begin{equation}
    \theta_{k} = \argmin_{\theta}\sum_{s_{t}^{i}, \mathcal{S}_{t+1}^{i}, a_{t}^{i}, i \in 1 \ldots k} || f_{\theta_k}( (s_{t}, g_{t+1}), (s_{t+1}, g_{t+1}) - a_{t} ) ||^{2}
\end{equation}
where $f$ is the SGD model \cite{robbins1951stochastic}. Moreover, Deep Q Network (DQN) \cite{mnih2015human} was applied in \cite{mahajan2023intent} to facilitate the intent sharing process of agents in highway merging scenarios. To reason about agents' obervations and control inputs given driver types, Chen and Chaudhari \cite{chen2021midas} employed Double DQN \cite{van2016deep} which consists of an online DQN and its periodic copy as the offline DQN. Additionally, the technique decouples the action selection from the action evaluation for reducing the issue of overestimation. To handle the large state space due to the action delay of agents in the autonomous navigation, Chen \textit{et al.} \cite{chen2020delay} and Fan \textit{et al.} \cite{fan2023switching} adopted Deep Deterministic Policy Gradient (DDPG) \cite{Lillicrap2015ContinuousCW} with the Actor-Critic (AC) algorithm \cite{peters2008natural} in CA-MAS. In these approaches, the actor samples actions based on agents' observations and policy gradient, and the critic evaluates each pair of observation and action and returns Q-value as feedback to improve the performance of the actor. For scalability and stability of the reasoning process, Everett \cite{everett2018motion} utilized the Asynchronous Advantage Actor-critic (A3C) method \cite{Babaeizadeh2016ReinforcementLT} which is the extension of AC and parallelizes agents experiences in either GPUs or CPUs. Furthermore, to reason about agents' driving actions and their instant response which are drawn from the two distributions (see also Section \ref{sec:camas_learn}), the Proximal Policy Optimization (PPO) \cite{schulman2017proximal} was employed in \cite{wu2023intent}.
}

\subsubsection{Predict and Act} \label{sec:camas_predict}
An agent projects scenarios or events that may occur in the near future. \blue{Once the projection is available, the agent performs its action to optimally achieve its goals, such as minimizing the estimation error or maximizing the cumulative rewards. However, in} CA-MAS with pre-defined rules, cases, or patterns \blue{\cite{beydoun2009security,castellano2014multi,ferrando2013context,feyzi2020model,fu2015adaptive,fuentes2006reputation,haiouni2019context,jakobson2006situation,kwon2004applying,olaru2013context,vladoiu2010learning,wan2008context,yousaf2023context}}, agents may lack prediction capability due to restricted conditions, weakening their ability to deal with uncertainties.
To address this challenge, agents are equipped with predictive models that estimate near-future events using weighting schemes \blue{\cite{jelen2022multi,lee2022density,riabchuk2022utility,twardowski2015iot,yan2018energy}}, probabilities \blue{\cite{katyal2020intent,tao2020dynamic,xie2021congestion,yusuf2022formalizing}}, or rewards \blue{\cite{chen2020delay,chen2021midas,everett2018motion,fan2023switching,huang2022intent,mahajan2023intent,nadi2017adaptive,qi2018intent,wu2023intent}}. 

\blue{
Predictive models with weighting schemes encompass regression, classification, and weighted prediction. In \cite{jelen2022multi}, agents employed a regression model to forecast charging station utilization and a classification model to determine parking lot availability for electric vehicle fleet routing. Similarly, in \cite{riabchuk2022utility}, agents utilized a classification model to assess user availability at specific times. This availability information, combined with data on user device usage, served as inputs for a regression model to predict energy consumption at those times. Furthermore, agents using weighted prediction models aim to forecast future data points based on weighted distributions \cite{lee2022density,yan2018energy} or aggregated weights \cite{twardowski2015iot}.
}

\blue{
Agents with probabilistic prediction models aim to project future stages based on either probabilistic event chains or probabilistic distributions. For trajectory planning, agents in \cite{katyal2020intent} and \cite{tao2020dynamic} predict the intents of other agents by analyzing the probability distribution of observations within autonomous navigation environments. Similarly, an agent in \cite{xie2021congestion} predicts future trajectories based on the probability distribution of its historical data. Furthermore, agents in \cite{yusuf2022formalizing} represent fire spread by projecting events through a probabilistic chain involving factors such as fuel types, fire presence, fuel conditions, and fire locations, utilizing a Bayesian Belief Network (BBN).
}

\blue{
An agent using reward-based models forecasts its sequential actions based on either the probability distribution \cite{chen2020delay,everett2018motion,mahajan2023intent,nadi2017adaptive,qi2018intent} or the weighted distribution \cite{chen2021midas,fan2023switching,huang2022intent,wu2023intent} of its observations and historical data. These actions are evaluated through policy and value functions within the reinforcement learning setting. The expected values of these functions are referred to as rewards. The agent's goal is to minimize predictions that result in negative rewards and maximize those that lead to positive rewards.
}

\section{Challenges and Future Directions} \label{sec:future}
CA-MAS is an emerging research topic. Although extensive efforts have been made on the integration between CAS and MAS in diverse problem domains, there are research directions worthy of future explorations.

Many context-aware multi-agent systems relax organizational constraints, causing two challenges. First, ineffective context sharing as one agent's irrelevant context introduces noise, reducing performance. \blue{Picard \textit{et al.} \cite{picard2009reorganisation} identified two perspectives in MAS: the agent-centered view and the organization-centered view. They also emphasized the connection between organizational awareness in MAS and the agents' ability to adapt and maintain autonomy. Furthermore, Esmaeili \textit{et al.} \cite{esmaeili2016impact} conducted an empirical study to demonstrate the effects of diversity on MAS performance. Similarly, Corkill \textit{et al.} \cite{corkill2016exploring} showed that agent organizations help to reduce the complexity of MAS and enhance system performance. Despite the existing literature on MAS, the impact of agent organizations has not been thoroughly explored in the context of CA-MAS. The second challenge is about security and privacy concerns. Due to the nature of CA-MAS, sensitive contextual information such as identity and location can be shared among agents. However, untrusted agents may exist without proper organizational structure that brings authority. These agents may threaten the system and grant unauthorized access that violates privacy regulations. Addressing this challenge requires investigating methodologies that incorporate organizational structures in CA-MAS to improve security and effective context sharing.}

The assumptions of consensus among agents have also been relaxed in existing CA-MAS. This leads to several issues in CA-MAS, such as uncertainty and ambiguity, where there is incomplete information and semantic mismatch between such information, as well as conflicts among agents \blue{\cite{kim2010output,qin2016recent,cho2018dynamics,li2019survey,amirkhani2022consensus}}. Addressing these issues enhances the robustness of agents in dealing with uncertainties and improves the effectiveness of learning strategies with joint information when solving specific tasks. However, solutions for these issues have not been actively discussed in the literature of CA-MAS. The application of consensus can be one potential solution. Research works such as those by \cite{yan2018energy} and \cite{xie2021congestion} have attempted leader-follower consensus to improve information sharing among agents. Besides this type of consensus, it is worth investigating techniques that utilize other types of consensus \cite{qin2016recent,li2019survey,amirkhani2022consensus} in CA-MAS for diverse problem domains. 

Agents' ontology provides the semantic meaning of information, supporting agents in comprehending situations for effective problem-solving. Many existing CA-MAS have employed ontology for both learning and reasoning tasks \blue{(see also Sections \ref{sec:camas_learn} and \ref{sec:reason})}; however, these CA-MAS heavily depend on pre-defined rules or patterns. This reliance affects the robustness of agents in dynamic environments where uncertain rules or patterns are constantly emerging. The integration between CA-MAS and \blue{goal-oriented reasoning frameworks (see also Section \ref{sec:camas_reason})} addresses such limitations. However, agents' ontology is not utilized in these systems. This may be due to two reasons: (1) the complexity of representing agents' ontology; and (2) the integration between those representations and \blue{goal-oriented} techniques. Just as Graph Neural Network (GNN) and its variants \cite{wu2020comprehensive} can potentially be utilized to address the first challenge; the Variational Auto-Encoder (VAE) architecture \cite{Kingma2013AutoEncodingVB,kipf2016semi} can be applied to \blue{produce the compatible learning representations for goal-oriented methods}. As this demonstrates the feasibility of the integration, it would be interesting to explore novel approaches that integrate agents' ontology into \blue{goal-oriented} CA-MAS to enhance the context comprehension of agents. 

\section{Conclusion} \label{sec:conclusion}

This survey provides a thorough overview of CA-MAS, representing the integration of CAS and MAS. We outline foundational knowledge to facilitate their integration and propose a general process for CA-MAS based on research across domains. Additionally, we highlight the challenges faced and suggest future research directions for these systems.

\section*{Acknowledgments}
This work is supported as part of the Higher Degree Research (HDR) program at the Applied Artificial Intelligence Institute $(A^2I^2)$, Deakin University.

\bibliographystyle{unsrt}  
\bibliography{references}

\end{document}